\documentclass[draft,10pt,noamsfonts,reqno]{amsart}
\usepackage{a4}
\usepackage{amssymb,upref}

\newcommand{\al}{\alpha}
\newcommand{\ga}{\gamma}
\newcommand{\eps}{\varepsilon}
\renewcommand{\th}{\vartheta}
\newcommand{\om}{\omega}

\newcommand{\fC}{\mathbb{C}}

\newcommand{\fN}{\mathbb{N}}
\newcommand{\fQ}{\mathbb{Q}}
\newcommand{\fR}{\mathbb{R}}
\newcommand{\fT}{\mathbb{T}}
\newcommand{\fZ}{\mathbb{Z}}
\newcommand{\eg}{\textit{e.g.~}}
\newcommand{\ie}{\textit{i.e.~}}
\newcommand{\der}{\text{\textup{d}}}
\newcommand{\tle}{\preccurlyeq}

\newcommand{\tl}{\prec}

\newcommand{\p}{\partial}
\newcommand{\elle}{\mathcal{L}}
\newcommand{\erre}{\mathcal{R}}
\newcommand{\enne}{\mathcal{N}}
\newcommand{\vu}{\mathcal{V}}
\newcommand{\effe}{\mathcal{F}}
\newcommand{\gi}{\mathcal{G}}
\newcommand{\ti}{\mathcal{T}}
\newcommand{\ee}{\mathcal{E}}
\newcommand{\aaa}{\mathcal{A}}
\newcommand{\bbb}{\mathcal{B}}
\newcommand{\pii}{\mathcal{P}}
\newcommand{\vres}{\mathbf{V}}
\newcommand{\wres}{\mathbf{W}}

\newcommand{\Val}{\operatorname{Val}}

\newtheorem*{thm}{Theorem}
\newtheorem{lem}{Lemma}

\theoremstyle{definition}

\theoremstyle{remark}
\newtheorem{rem}{Remark}
\newtheorem*{defin}{Definition}

\newtheoremstyle{case}{5pt}{5pt}{}{}{\scshape}{ }{ }%
{\thmnote{[#3]}}
\theoremstyle{case}
\newtheorem*{case}{Case}

\numberwithin{equation}{section}

\begin{document}

\title{Bryuno Function and the Standard Map}

\author{Alberto Berretti}
\address{Alberto Berretti\\ Dipartimento di Matematica\\ II Universit\`{a}
di
Roma (Tor Vergata)\\ Via della Ricerca Scientifica, 00133 Roma, Italy and
Istituto Na\-zio\-na\-le di Fisica Nucleare,
Sez. Tor Vergata}
\email{{\tt berretti@roma2.infn.it}}

\author{Guido Gentile}
\address{Guido Gentile\\ Dipartimento di Matematica\\ Universit\`{a}
di Roma 3\\ Largo S. Leonardo Murialdo 1, 00146 Roma, Italy}
\email{{\tt gentile@matrm3.mat.uniroma3.it}}

\begin{abstract}
    For the standard map the homotopically 
    non-trivial invariant cur\-ves of rotation number $\om$ satisfying 
    the Bryuno condition are shown to be analytic in the
    perturbative parameter $\eps$, provided $|\eps|$ is small enough.
    The radius of convergence $\rho(\om)$ of the Lindstedt
    series -- sometimes called \emph{critical function}
    of the standard map -- is studied and the 
    relation with the Bryuno function $B(\om)$ is
    derived: the quantity $|\log\rho(\om) + 2 B(\om)|$ is proved
    to be bounded uniformily in $\om$.
\end{abstract}

\maketitle

\section{Introduction}
\label{sect:intro}

\noindent
We continue the study, started in \cite{BG1}, of the radius of 
convergence of the Lindstedt series for the standard map, for rotation 
numbers close to rational values. We consider real rotation numbers 
$\om$ satisfying the Bryuno condition (see below), and study
how the corresponding radius of convergence depends
on the Bryuno function $B(\om)$, introduced by Yoccoz in \cite{Y}. 

The standard map is a discrete time, one-dimensional dynamical system 
generated by the iteration of the area-preserving -- symplectic -- map 
of the cylinder into itself $T_{\eps}: \fT\times\fR \mapsto 
\fT\times\fR$, given by:
\begin{equation}
    T_{\eps}: \; \begin{cases}
    x' = x + y + \eps\sin x,\\
    y' = y + \eps\sin x.
    \end{cases}
    \label{eq:s1.sm}
\end{equation}
Given a real \emph{rotation number} $\om \in [0,1)$, we can look for 
(homotopically non-trivial) invariant curves described parametrically 
by:
\begin{equation}
    \begin{cases}
	x = \al + u(\al,\eps;\om),\\
	y = \al + u(\al,\eps;\om) - u(\al - 2\pi\om,\eps;\om),
    \end{cases}
    \label{eq:s1.param}
\end{equation}
such that the dynamics induced in the variable $\al$ is given by 
rotations by $\om$:
\begin{equation}
    \al' = \al + 2\pi\om.
    \label{eq:s1.lindyn}
\end{equation}
For irrational rotation numbers $\om$, by imposing that the average 
of $u$ over $\al$ is $0$, the (formal) conjugating function $u$ is 
unique and odd in $\al$, and has a formal expansion -- known as 
Lindstedt series -- of the form:
\begin{equation}
    u(\al,\eps) = \sum_{\nu \in \fZ}u_{\nu}(\eps)e^{i\nu\al} = 
    	\sum_{k \geq 1} u^{(k)}(\al)\eps^{k} = 
	\sum_{k \geq 1}\sum_{\nu \in \fZ}u^{(k)}_{\nu}e^{i\nu\al}\eps^{k};
    \label{eq:s1.lind}
\end{equation}
the coefficients $u^{(k)}_{\nu}$ can be expressed graphically in terms 
of sums over \emph{trees} as explained shortly (see also \cite{BG1} 
and references quoted therein). The \emph{radius of convergence} of 
the series \eqref{eq:s1.lind}, called sometimes the \emph{critical 
function} of the standard map, is defined as:
\begin{equation}
    \rho(\om) = \inf_{\al\in\fT}\Bigl(\limsup_{k \rightarrow \infty}
    	\bigl|u^{(k)}(\al)\bigr|^{1/k}\Bigr)^{-1}.
    \label{eq:s1.raddef}
\end{equation}
Given $\om$, let $\{p_{n}/q_{n}\}$ be the sequence of 
\emph{convergents} defined by the standard continued fraction 
expansion of $\om$, and let:
\begin{equation}
    B_{1}(\om) = \sum_{n=0}^{\infty}\frac{\log q_{n+1}}{q_{n}}.
    \label{eq:s1.bry1}
\end{equation}
The irrational number $\om \in [0,1)$ \emph{satisfies the Bryuno 
condition} if $B_{1}(\om) < \infty$; we also say that in this case 
$\om$ is a \emph{Bryuno number}.  After Yoccoz \cite{Y}, we define on 
the irrational numbers the Bryuno function $B(\om)$ by the functional 
equation:
\begin{equation}
    \begin{cases}
	B(\om) = -\log\om + \om B(\om^{-1}) \quad 
	    \text{for $\om \in (0,\frac{1}{2})$ and irrational},\\
	B(\om+1) = B(-\om) = B(\om).
    \end{cases}
    \label{eq:s1.bry}
\end{equation}
It can be proved that such functional equation has a unique solution 
in $L_{p}$, $p \geq 1$; moreover $B(\om)$ is related to the series 
$B_{1}(\om)$ by the inequality:
\begin{equation}
    \bigl|B(\om) - B_{1}(\om)\bigr| < C_{1} ,
    \label{eq:s1.bryrel}
\end{equation}
for some constant $C_1$.
See \cite{Y} and \cite{MMY} for the proofs of these statements. 

We prove the following theorem.

\begin{thm}
    Consider the standard map \eqref{eq:s1.sm} and let $\om$ be an 
    irrational number, $\om \in [0,1)$, satisfying the Bryuno condition. 
    Then the radius of convergence \eqref{eq:s1.raddef} satisfies the 
    bound:
    \begin{equation}
        |\log\rho(\om) + 2 B(\om)| \leq C_{0},
        \label{eq:s1.bound}
    \end{equation}
    where $C_{0}$ is a constant independend on $\om$.
\end{thm}

An analogous result was proved by Davie \cite{D1} for the 
semistandard map (where the nonlinear term $\sin x$ in \eqref{eq:s1.sm} 
is replaced by $e^{ix}$); in the same paper it was also shown that 
the upper bound in \eqref{eq:s1.bound} holds:
\begin{equation}
    \log\rho(\om) + 2 B(\om) < C_{2},
    \label{eq:s1.ubound}
\end{equation}
for some constant $C_{2}$.
In ref.  \cite{D2} it was proved, by ``phase space renormalization'' 
arguments, that $\forall \eta > 0$ $\exists C_{3}$, depending on 
$\eta$, such that:
\begin{equation}
    \log\rho(\om) + (2+\eta) B(\om) > C_{3}.
    \label{eq:s1.dlbound}
\end{equation}
So our theorem improves the result of \cite{D2} (using also a 
different, direct technique, taken from \cite{GM}
-- and inspired to the works \cite{E} and \cite{G} --,
in some sense more elementary 
than the one of \cite{D2}) and proves for the standard map the 
conjecture first stated in \cite{MS}. 

The paper is organized as follows. In sect. \ref{sect:trees}
we introduce the formalism and give the scheme of the
proof of the theorem, elucidating the major difficulties,
due to the accumulation of small divisors in the
Lindstedt series, and showing that, in absence of
such a phenomenon, the proof could be carried out
by a detailed analysis of the single terms of the series.
In sect. \ref{sect:canc1} and \ref{sect:canc3}, we shall see
how to handle the small divisors problem, by showing that there are
cancellation mechanisms, operating to all perturbative orders between
different terms of the Lindstedt series, which assure its convergence.
Finally sect. \ref{sect:proof4} and \ref{sect:proof6} deal with
the proof of the main technical lemmata used in the proof of the theorem.

\vspace{1cm}

\section{Formalism: trees, clusters and resonances}
\label{sect:trees}

\noindent
As in \cite{BG1}, we can express graphically the coefficients 
$u^{(k)}_{\nu}$ in \eqref{eq:s1.lind} in terms of \emph{trees}.  We 
shall only recall the definitions used in this paper and set up the 
notations, leaving the full details of the tree expansion for our 
problem to \cite{BG1} and the references quoted therein.

A tree $\th$ consists of a family of lines arranged to connect a 
partially ordered set of points -- nodes --, with the lower nodes to the 
right.  All the lines have two nodes at their extremes, except the 
highest which has only one node, the \emph{last node} $u_{0}$ of the 
tree; the other extreme $r$ will be called the \emph{root} of the tree 
and it will not be regarded as a node.

We denote by $\tle$ the partial ordering relation between nodes 
defined as follows: given two nodes $u$, $v$, we say that $u \tle v$ 
if $u$ is along the path of lines connecting $v$ to the root $r$ of 
the tree -- they could coincide: we say that $u \tl w$ if they do 
not. So our trees are ``rooted trees'', following the terminology of 
\cite{HP}.

We assign to each line $\ell$ joining two nodes $u$ and $u'$ an 
``arrow'' pointing from the highest to the lowest node 
according to the order relation just defined; if $u' \tl u$, we say 
that the line $\ell$ exists from $u$ and enters $u'$.  We write 
$u'_{0} = r$ even if, strictly speaking, $r$ is not considered a node.  
For each node $u$ there is a unique exiting line, and $m_{u} \geq 0$ 
entering lines; as there is a one-to-one correspondence between lines 
and nodes, we can associate to each node $u$ the line $\ell_{u}$ 
exiting from it.  The line $\ell_{u_{0}}$ exiting the last node $u_{0}$ 
will be called the \emph{root line}.  Note that each line $\ell$ can 
be considered the root line of the subtree consisting of the nodes 
satisfying $v \tle u$, and $u'$ will be the root of such tree.  The 
\emph{order} $k$ of the tree is defined as the number of its nodes.

To each node $u \in \th$ we associate a \emph{mode label} $\nu_{u} = 
\pm 1$, and define the \emph{momentum} flowing through the line 
$\ell_{u}$ as:
\begin{equation}
	\nu_{\ell_{u}} = \sum_{w \tle u} \nu_{w}, \quad \nu_{w} = \pm 1;
	\label{eq:s2.momdef}
\end{equation}
note that no line can have zero momentum, as $u^{(k)}_{0} = 0$. 

While in \cite{BG1} we could get along considering only two 
``scales'', we need a full multiscale decomposition of the momenta 
associated to each line. 

Given a rotation number $\om \in [0,1) \backslash \fQ$, let 
$\{p_{n}/q_{n}\}$ be the sequence of convergents coming from the 
standard continued fraction expansion of $\om$. For $x \in \fR$, let:
\begin{equation}
	||x|| = \inf_{\nu \in \fZ}|x - p|
	\label{eq:s2.distdef}
\end{equation}
be the distance of $x$ from the nearest integer. Let now:
\begin{equation}
	\ga(\nu) = 2(\cos 2\pi\om\nu - 1);
	\label{eq:s2.gadef}
\end{equation}
then we have the estimate:
\begin{equation}
	|\ga(\nu)| = 2|\cos 2\pi\om\nu - 1| \geq \Gamma||\om\nu||^2,
	\label{eq:s2.gaest}
\end{equation}
for some constant $\Gamma$. 

We introduce a $C^{\infty}$ partition of unity in the following way. 
Let $\chi(x)$ a $C^{\infty}$, non-increasing, compact-support function 
defined on $\fR^{+}$, such that:
\begin{equation}
	\chi(x) = \begin{cases} 
		1&\quad \text{for $x \leq 1$},\\
		0&\quad \text{for $x \geq 2$},
	\end{cases}
	\label{eq:s2.chidef}
\end{equation}
and define for each $n \in \fN$:
\begin{equation}
	\begin{cases}
		\chi_{0}(x) = 1 - \chi(96q_{0}x),\\
		\chi_{n}(x) = \chi(96q_{n}x) - \chi(96q_{n+1}x),
			\quad \text{for $n \geq 1$}.
	\end{cases}
	\label{eq:s2.chindef}
\end{equation}
Then for each line $\ell$ set:
\begin{equation}
	g(\nu_{\ell}) \equiv \frac{1}{\ga(\nu_{\ell})} = 
		\sum_{n=0}^{\infty}
		\frac{\chi_{n}(||\om\nu_{\ell}||)}{\ga(\nu_{\ell})}
		\equiv \sum_{n=0}^{\infty}g_{n}(\nu_{\ell}),
	\label{eq:s2.multiscale}
\end{equation}
and call $g_{n}(\nu_{\ell})$ the \emph{propagator on scale $n$}. 

Given a tree $\th$, we can associate to each line $\ell$ of $\th$ a
scale label $n_{\ell}$, using the multiscale decomposition 
\eqref{eq:s2.multiscale} and singling out the summands with
$n = n_{\ell}$. We shall call $n_{\ell}$ the \emph{scale label} of
the line $\ell$, and we shall say also that the line $\ell$ is
\emph{on scale $n_{\ell}$}. 

\begin{rem}\label{rem:s2.rem1}
	Given a value $\nu_{\ell}$ there can be at most two possible -- 
	consecutive -- values of $n$ such that the corresponding 
	$\chi_{n}(||\om\nu_{\ell}||)$ are not vanishing.  This means that 
	at most only two summands of the infinite series 
	\eqref{eq:s2.multiscale} really appear; nevertheless keeping all 
	terms is more convenient, in order to have a label to
	characterize the ``size'' of the ``propagators'' $g(\nu_{\ell})$.
\end{rem}

\begin{rem}\label{rem:s2.rem2}
	Note that if a line $\ell$ has momentum $\nu_{\ell}$ and scale
	$n_{\ell}$, then:
	\begin{equation}
		\frac{1}{96q_{n_{\ell}+1}} \leq ||\om\nu_{\ell}|| \leq 
			\frac{1}{48q_{n_{\ell}}},
		\label{eq:s2.distest}
	\end{equation}
provided that one has $\chi_{n_{\ell}}(||\om\nu_{\ell}||) \neq 0$.
\end{rem}

A group $\gi$ of tranformations acts on the trees, generated by the 
permutations of all the subtrees emerging from each node with at least 
one entering line: $\gi$ is therefore a cartesian product of copies of 
the symmetric groups of various orders. Two trees that can be 
transformed into each other by the action of the group $\gi$ are 
considered identical. 

Denote by $\ti_{\nu,k}$ the set of trees, with
nonvanishing value, of order $k$ and total 
momentum $\nu_{\ell_{u_{0}}} = \nu$, if $u_{0}$ is the last node of 
the tree.  The number of elements in $\ti_{\nu,k}$ is bounded by 
$2^{k} \cdot 2^{k} \cdot 2^{2k} = 2^{4k}$: the number of 
semitopological trees (see \cite{BG1}) of order $k$ is bounded by 
$2^{2k}$,\footnote{The number of semitopological trees can be bounded 
by the number of one-dimensional random walks with $2k-1$ steps.} and 
there are two possible values for the mode label of each node and two 
possible values for the scale label of each line. 

Then, as in \cite{BG1} -- to which we refer for more details and 
figures -- one finds:
\begin{equation}
	u^{(k)}_{\nu} = \frac{1}{2^{k}}\sum_{\th \in \ti_{\nu,k}}\Val(\th),
	\quad 
	\Val(\th) = -i\Biggl[
			\prod_{u \in \th}\frac{\nu_{u}^{m_{u}+1}}{m_{u}!}
		\Biggr] \Biggl[
			\prod_{\ell \in \th}g_{n_{\ell}}(\nu_{\ell})
		\Biggr];
	\label{eq:s2.treeexp}
\end{equation}
the factors $g_{n_{\ell}}(\nu_{\ell})$ above are called \emph{propagators}
of 
\emph{small divisors} on scale $n_{\ell}$, and the quantity $\Val(\th)$ 
will be called the \emph{value} of the tree $\th$. 

We define now the main combinatorial tools.

\begin{defin}[Cluster]\label{def:s2.cluster}
	Given a tree $\th$, a cluster $T$ of $\th$ on scale $n$ is a maximal 
	connected set of lines of lines on scale $\leq n$ with at least one 
	line on scale $n$. We shall say that such lines are \emph{internal} 
	to $T$, and write $\ell \in T$ for an internal line $T$. A 
	node $u$ is called \emph{internal} to $T$, and we write $u 
	\in T$, if at least one of its entering lines or exiting line is in 
	$T$. Each cluster has an arbitrary number $m_{T} \geq 0$ of entering 
	lines but only one exiting line; we shall call \emph{external} to 
	$T$ the lines entering or exiting $T$ (which are all on scale $> 
	n$). We shall denote with $n_{T}$ the scale of the cluster $T$,
	with $n^{i}_{T}$ the minimum of the scales of 
	the lines entering $T$, with $n^{o}_{T}$ the scale of the line 
	exiting $T$ and with $k_{T}$ the number of nodes in $T$. 
\end{defin}

Note that, despite the name, not all lines outside $T$ are ``external'' 
to it: only those lines outside $T$ which enter or exit $T$ are 
external to it. On the contrary a line inside $T$ is said to be
``internal'' to it. The use of such a terminology is inherited from
Quantum Field Theory.

\begin{defin}[Resonance]\label{def:s2.resonance}
	Given a tree $\th$, a cluster $V$ of $\th$
	will be called a \emph{resonance} with \emph{resonance-scale}
	$n=n^{R}_{V} \equiv \min\{n^{i}_{V},n^{o}_{V}\}$, if:
\begin{enumerate}
	\item\label{enum:s2.res1}  the sum of the mode labels of its nodes is 
	$0$:
\begin{equation}
	\nu_{V} \equiv \sum_{u \in V}\nu_{u} = 0;
	\label{eq:s2.sumres}
\end{equation}

	\item\label{enum:s2.res2}  all the lines entering $V$ are on the 
	same scale except at most one, which can be on a higher scale;

	\item\label{enum:s2.res3}  $n^{i}_{V} \leq n^{o}_{V}$ if $m_{V} \geq 
	2$, and $|n^{i}_{V} - n^{o}_{V}| \leq 1$ for $m_{V} = 1$;

	\item\label{enum:s2.res4} $k_T<q_{n}$;

	\item\label{enum:s2.res5} $m_V=1$ if $q_{n+1}\leq 4q_n$;

	\item\label{enum:s2.res6} if $q_{n+1}> 4q_{n}$
	and $m_V\geq 2$, denoting by $k_0$
	the sum of the orders of the subtrees of order
	$<q_{n+1}/4$ entering $V$, either

	\begin{enumerate}
		\item\label{enum:s2.res61}
		there is a only one subtree of order
		$k_1 \geq q_{n+1}/4$ entering $V$ and
		$k_1+k_0+k_T\geq q_{n+1}/4$, $k_0<q_{n+1}/8$, or
		
		\item\label{enum:s2.res62}
		there is no such subtree and $k_0+k_T<q_{n+1}/4$.
	\end{enumerate}
\end{enumerate}
\end{defin}

\begin{rem}
	Note that for any resonance $V$ one has $n^{R}_{V} \geq n_{V} + 1$, 
	if $n_{V}$ is the scale of the resonance $V$ as a cluster.
	As in \cite{GG} we use the notation with a hyphen for the
	resonance-scale to avoid confusion between $n_V^R$ and $n_V$.
\end{rem}

\begin{rem}
	One would be tempted to give a simpler definition of resonance (for 
	instance, by imposing only condition \ref{enum:s2.res1} to the 
	cluster $V$). This temptation should be resisted, as it would make 
	impossible to exploit the cancellations leading to the improvement of 
	the bound discussed at the end of this section (in fact,
	no relation would continue to subsist between momenta
	and scale labels and factorials 
	would arise from counting the summands generated by the 
	renormalization procedure described in sect. \ref{sect:canc3}).
	On the other hand we shall see in sect. \ref{sect:proof4}
	that no problems should arise if no resonances -- exactly
	as they defined above -- could appear.
\end{rem}


In the following we shall need to introduce trees in which it can happen
that a line $\ell$ is on a scale $n_{\ell}$ and yet its momentum
does not satisfy \eqref{eq:s2.distest}. The value of any such tree $\th$
is vanishing as $\chi_{n_{\ell}}(||\om\nu_{\ell}||)=0$; nevertheless
it will be useful to write $\Val(\th)$ as sum of two (possibly)
nonvanishing terms: one of them will be used to cancel terms arising
from other tree values, so it will disappears, while
the other one is left and has to be bounded. This means that
we shall have to deal with trees in which there are lines $\ell$
with momentum $\nu_{\ell}$ and scale $n_{\ell}$ which do not satisfy
\eqref{eq:s2.distest}. What will be shown to hold is that for
such lines a bound similar to \eqref{eq:s2.distest}, though weaker,
still holds; more precisely, a line $\ell$ with momentum $\nu_{\ell}$
will have only scales $n_{\ell}$ such that
\begin{equation}
	\frac{1}{768q_{n_{\ell}+1}} \leq ||\om\nu_{\ell}|| \leq 
		\frac{1}{8q_{n_{\ell}}},
	\label{eq:s2.distest1elle}
\end{equation}
and, for fixed $\nu_{\ell}$,
the number of possible scales to associate to $\ell$
is bounded by an absolute constant.

As \eqref{eq:s2.distest1elle} is implied by \eqref{eq:s2.distest},
even for trees with nonvanishing value we shall use that if
a line is on scale $n_{\ell}$ then \eqref{eq:s2.distest1elle} holds.

Then, if $N_{n}(\th)$, $n \in \fN$, denotes the number of lines on scale 
$n$ in $\th$, we have trivially for a given tree $\th$ the bound:
\begin{equation}
	|\Val(\th)| \leq D_{1}^{k}\prod_{n=0}^{\infty}
		\bigl(768q_{n+1}\bigr)^{2N_{n}(\th)},
	\label{eq:s2.trivbound}
\end{equation}
for some constant $D_{1}$ (actually $D_{1} = 1/\Gamma$; see 
\eqref{eq:s2.gaest}, \eqref{eq:s2.treeexp} and 
\eqref{eq:s2.distest1elle}).

Given a tree $\th$, let us denote with $N^{R}_{n}(\th)$ the number of 
resonances with res\-on\-ance-sca\-le $n$ and by $P_{n}(\th)$ the number of

resonances on scale $n$. Of course $N^{R}_{0} = 0$. 

\begin{rem}\label{rem:s2.remark5}
	Note that the number $N^{R}_{n}(\th)$ of resonances with
	resonance-scale $n$
	can be counted by counting \emph{the number of lines exiting 
	resonances with resonance-scale $n$}; analogously $P_{n}(\th)$ can 
	be counted by counting the number of lines exiting resonances on 
	scale $n$. Such counts will be performed in sect. \ref{sect:proof4}.
\end{rem}

The following simple lemmata contain all the arithmetic we shall need, 
and are basically adapted from \cite{D1}. 

\begin{lem}[Davie's lemma]\label{lem:s2.davie}
	Given $\nu\in\fZ$ such that
	$||\om\nu|| \leq 1/4q_{n}$, then
	\begin{enumerate}
		\item \label{enum:s2.d1} either $\nu=0$ or $|\nu| \geq q_{n}$,

		\item \label{enum:s2.d2} either $|\nu| \geq q_{n+1}/4$ or
		$\nu = sq_{n}$ for some integer $s$.
	\end{enumerate}
\end{lem}

\begin{lem}\label{lem:s2.smalltree}
	If a tree $\th$ has $k < q_{n}$ nodes, then $N_{n}(\th) = 0$ and 
	$P_{n-1}(\th) = 0$.
\end{lem}

\begin{lem}\label{lem:s2.resto}
	For any irrational number $\om\in[0,1)$:
\begin{equation}
	\sum_{n=0}^{\infty}\frac{\log q_{n}}{q_{n}} \leq D_{2},
	\label{eq:s2.resto}
\end{equation}
	for a constant $D_{2}$; here $q_{n}$ are the denominators of the 
	convergents of $\om$. 
\end{lem}

\begin{lem}\label{lem:s2.scale}
	Given a momentum $\nu$ such that
	\begin{equation}
		\frac{1}{768q_{n+1}} \leq ||\om\nu|| \leq 
			\frac{1}{8q_{n}},
		\label{eq:s2.distest1}
	\end{equation}
	then one can have $\chi_{n'}(||\om\nu||)\neq 0$ only
	for $n'$ such that $n-8 \leq n' \leq n+8$.
\end{lem}

\begin{proof}[Proof of lemma \ref{lem:s2.davie}]
If $\{q_{n}\}$ are the denominators of the convergents of $\om$, 
then (see \eg \cite{Schmidt}, Ch. 1, \S  3):
\begin{equation}
	\frac{1}{2q_{n+1}} < ||\om q_{n}|| < \frac{1}{q_{n+1}},
	\label{eq:s2.tmp1}
\end{equation}
and:
\begin{equation}
	\forall |\nu|< q_{n+1}, |\nu| \neq q_{n}: \quad 
		||\om\nu|| > ||\om q_{n}||.
	\label{eq:s2.tmp2}
\end{equation}
To prove \ref{enum:s2.d1} note that if $\nu=0$ nothing has to be
proved: so we assume $\nu\neq 0$.
If $|\nu|<q_n$, by \eqref{eq:s2.tmp2} and \eqref{eq:s2.tmp1}, 
$||\om\nu|| \geq ||\om q_{n-1}|| > 1/2q_{n}$, so that $||\om\nu|| < 
1/4q_{n}$ implies $|\nu| \geq q_{n}$,
proving the first assertion of lemma \ref{lem:s2.davie}. 

To prove \ref{enum:s2.d2}, again if $\nu=0$ nothing has to be
proved (and $s=0$): so we assume $\nu\neq 0$, and proceed
by \textit{reductio ad absurdum}. If 
$0 < \nu < q_{n+1}/4$ and there does not exist any $s \in \fZ$ such
that $\nu = sq_{n}$, then one has $\nu = mq_{n}+r$, with $0 < r < 
q_{n}$ and $m < q_{n+1}/4q_{n}$; then, by \eqref{eq:s2.tmp1},
$||\om m q_{n}|| \leq m||\om q_{n}||< m/q_{n+1} < 1/4q_{n}$, and, by
\eqref{eq:s2.tmp2}, $||\om r|| \geq ||\om q_{n-1}||
> 1/2q_{n}$, as $r\neq 0$; so $||\om\nu|| \geq ||\om r||-
||\om m q_{n}|| > 1/4q_{n}$.
The case $0 > \nu > -q_{n+1}/4$ is identical as $||\cdot||$ is even.
\end{proof}

\begin{proof}[Proof of lemma \ref{lem:s2.smalltree}]
If $k < q_{n}$, then for any $\ell \in \th$ one has $|\nu_{\ell}| \leq 
k < q_{n}$, so that, by \eqref{eq:s2.tmp1} and \eqref{eq:s2.tmp2}, 
$||\om\nu_{\ell}|| \geq ||\om q_{n-1}|| > 1/2q_{n}$, hence $n_{\ell} < 
n$ and so $N_{n}(\th) = 0$.  If there are no lines on scale $n$, it is 
impossible to form a cluster on scale $n-1$, \textit{a fortiori} a 
resonance.
\end{proof}

\begin{proof}[Proof of lemma \ref{lem:s2.resto}]
The denominators of the
convergents $\{q_{n}\}$ of $\om$ satisfy $q_{0}=1$, $q_{1} \geq 1$ 
and $q_{n} \geq 2q_{n-2}$ for any $n \geq 2$. So we can write:
\begin{equation}
	\sum_{n=0}^{\infty}\frac{\log q_{n}}{q_{n}} = 
		\sum_{n=0}^{\infty}\frac{\log q_{2n}}{q_{2n}} + 
		\sum_{n=0}^{\infty}\frac{\log q_{2n+1}}{q_{2n+1}};
	\label{eq:s2.restotemp}
\end{equation}
using the fact that, for $x \geq e$, $x^{-1}\log x$ is decreasing, we 
obtain easily:
\begin{equation}
	\sum_{n=0}^{\infty}\frac{\log q_{n}}{q_{n}} \leq 
		3\max_{x \geq 1}\Biggl\{\frac{\log x}{x}\Biggr\} + 
		2\log 2 \sum_{k=2}^{\infty}\frac{k}{2^{k}} = 
		3(e^{-1} + \log 2) \equiv D_{2},
	\label{eq:s2.estlemresto}
\end{equation}
which also gives an explicit value for the constant $D_{2}$.
\end{proof}

\begin{proof}[Proof of lemma \ref{lem:s2.scale}]
Simply use that $q_{n+1}\geq q_n$ and $q_{n+2}\geq 2q_n$ for all $n\geq 0$,
to deduce that $1/48q_{n+9}<1/768q_{n+1}$ and $1/96q_{n-8}>1/8q_{n}$.
\end{proof}

The following ``counting'' lemma is the main result stated in this 
section, and it can be considered an adaption and extension of lemma 
$2.3$ in \cite{D1}. We postpone its proof to sect. \ref{sect:proof4}.

\begin{lem}\label{lem:s2.count}
	Given a tree $\th$, let $M_{n}(\th) = N_{n}(\th) + P_{n}(\th)$. Then:
\begin{equation}
	M_{n}(\th) \leq \frac{k}{q_{n}} + \frac{8k}{q_{n+1}} + N^{R}_{n}(\th),
	\label{eq:s2.count}
\end{equation}
where $k$ is the order of $\th$. 
\end{lem}

Therefore we can rewrite the bound \eqref{eq:s2.trivbound} on the 
tree value as:
\begin{equation}
	\begin{split}
		|\Val(\th)| & \leq D_{1}^{k}\prod_{n=0}^{\infty}
		\bigl(768q_{n+1}\bigr)^{2(M_{n}(\th) - P_{n}(\th))} \\
			& \leq D_{1}^{k}\prod_{n=0}^{\infty}
		\bigl(768q_{n+1}\bigr)^{2(k/q_{n}+8k/q_{n+1}+
			N^{R}_{n}(\th)-P_{n}(\th))}.
	\end{split}
	\label{eq:s2.trivbound2}
\end{equation}

Note that at this point it would be very easy to prove the lower bound 
in \eqref{eq:s1.bound} \emph{for the semistandard map} and, by simple
modifications of the same scheme, \emph{for Siegel problem}, since in these
cases no 
resonances appear.  On the contrary, in the more difficult case of the 
standard map we lack, for the moment, a control on the number 
$N^{R}_{n}(\th)$ of resonances in $\th$ with resonance-scale $n$. 

In sect. \ref{sect:canc1} and \ref{sect:canc3} we shall see how to 
improve the bound \emph{on the sum} over the trees of fixed order and
total momentum, in order to prove the theorem stated in sect.
\ref{sect:intro}. We postpone to forthcoming sections the proofs, 
limiting ourselves here to a heuristic discussion in order to give an 
idea of the structure of the proof.

We perform a suitable resummation -- described in sect.  
\ref{sect:canc1} and \ref{sect:canc3} -- whose consequence is that, for 
each resonance $V$, \emph{it is as if one of the external lines on scale
$n^{R}_{V}$ contributed $\bigl(768q_{n_{V}+1}\bigr)^{2}$ instead of
$\bigl(768q_{n^{R}_{V}+1}\bigr)^{2}$}. To obtain such a result, we shall
perform on trees transformations which will lead to the introduction of
new trees: so we extend $\ti_{\nu,k}$ to a larger set
$\ti_{\nu,k}^*$. However we shall prove that the value of each single
tree in $\ti_{\nu,k}^*$ still admits the bound \eqref{eq:s2.trivbound2}
-- even if, unlike the values of the trees in $\ti_{\nu,k}$,
it fails to satisfy the same bound with $768$ replaced with $96$ --
and the number of elements in $\ti_{\nu,k}^*$ is bounded by a constant
to the power $k$ (\ie no bad counting factors, like factorials, appear).
Then we obtain, \emph{for the sum of the resummed trees},
a bound of the form \eqref{eq:s2.trivbound2} with:
\begin{equation*}
	\prod_{n=0}^{\infty}\bigl(768q_{n+1}\bigr)^{2N^{R}_{n}(\th)}
\end{equation*}
replaced with:
\begin{equation*}
	D_{3}^{k}\prod_{n=0}^{\infty}\bigl(768q_{n+1}\bigr)^{2P_{n}(\th)},
\end{equation*}
for some constant $D_{3}$. By using that the number of trees in 
$\ti_{\nu,k}^*$ will be shown to be bounded by a constant to the
power $k$, we obtain, for some constants $D_{4}$, $D_{5}$:
\begin{equation}
	\begin{split}
	        |u^{(k)}(\al)| & \leq
		\biggl|\sum_{|\nu|\leq k}\sum_{\th\in\ti_{\nu,k}}
			\Val(\th)\biggr| \leq
	\biggl|\sum_{|\nu|\leq k}\sum_{\th\in\ti_{\nu,k}^*}
			\Val(\th)\biggr| \\
		& \leq D_{4}^{k}\prod_{n=0}^{\infty}
			\bigl(768q_{n+1}\bigr)^{2k/q_{n}+16k/q_{n+1}} \\
		& \leq D_{5}^{k} \exp\biggl[2k\sum_{n=0}^{\infty}\biggl(
			\frac{\log q_{n+1}}{q_{n}} +
			\frac{8\log q_{n+1}}{q_{n+1}}
			\biggr)\biggr],
	\end{split}
	\label{eq:s2.trivbound3}
\end{equation}
which, by making use of lemma \ref{lem:s2.resto}, gives:
\begin{equation}
	\log\rho(\om) + 2B_{1}(\om) \geq -16D_{2} - \log D_{5}.
	\label{eq:s2.trivbound4}
\end{equation}
By making rigorous the above discussion in sect. \ref{sect:canc1}
and \ref{sect:canc3}, we shall complete the proof of the
theorem, since the bound from above was already proved in \cite{D1}.

\vspace{1cm}

\section{Renormalization of resonances: set-up and the first step}
\label{sect:canc1}

\noindent
Given a tree $\th$, let us consider maximal resonances, \ie resonances 
\emph{not} contained in any larger resonance; let us call them 
\emph{first generation resonances}.  Inside the first generation 
resonances let us consider the ``next maximal'' resonances, \ie the 
resonances not contained in any larger resonance except first 
generation resonances, and let us call them \emph{second generation 
resonances}.  We can define in this way \emph{$j$-th generation 
resonances}, for $j\ge 2$, as resonances which are maximal
within $(j-1)$-th generation resonances. 

Let $\vres$ be the set of all resonances of a tree $\th$, and 
$\vres_{j}$ the set of all resonances of $j$-th generation,
with $j=1,\ldots,G$, for some integer $G$, depending on $\th$.

Given a tree $\th$ and a resonance $V\in\vres_{j}$
with $m_{V}$ entering lines, define $V_{0}$ as the set of nodes
and lines internal to $V$ and outside any resonances contained in $V$.
Let $L_{V}=\{\ell_{1},\ldots,\ell_{m_{V}}\}$ be the set of
entering lines of $V$; we define $L_{V}^{R}$ as the subset of
the lines in $L_V$ which enter some resonances of higher generation
contained inside $V$ and $L_{V}^{0}=L_{V}\setminus L_{V}^{R}$
as the subset of lines in $L_{V}$ which enter nodes in $V_{0}$.

For any line $\ell_{m}\in L_{V}^{R}$, let $V(\ell_{m})$ be
the minimal resonance containing the node which the line $\ell_{m}$
enters (\emph{i.e.} the highest generation resonance containing such
a node) and $V_{0}(\ell_{m})$ the set of nodes and lines internal to
$V(\ell_{m})$ and outside resonances contained in $V(\ell_{m})$. Define:
\begin{equation}
	\tilde\vres (V) = \{ \tilde V \subset V\,: \;
	\tilde V = V(\ell_{m}) \text{ for some }
	\ell_{m}\in L_{V}^{R} \} .
	\label{eq:s3.tilderes}
\end{equation}

Call $m_{V_{0}}$ the number of lines in $L_{V}^{0}$.
The number of lines in $L_{V}^{R}$ entering the
same resonance $\tilde V\in \tilde\vres(V)$ is not arbitrary:
it is always 1, as it is shown by the following lemma.

\begin{lem}\label{lem:s4.lemma7}
For $j \geq 1$, given a resonance $W \in \vres_{j+1}$ contained inside 
a resonance $V \in \vres_{j}$, only one among the entering lines $W$ 
can also enter $V$. 
\end{lem}

\begin{proof}
By item \ref{enum:s2.res3} of the definition of resonance one has
$n_{W}^{R}\leq n_{V}$, otherwise $V$ would be a cluster on scale
$< n^{R}_{W}$, so that all the lines external to $W$ 
would be also external to $V$ and $V=W$, while we assumed that $V 
\varsubsetneq W$. Then if a line $\ell$ enter both $V$ and $W$,
one must have $n_{\ell}>n_{W}^{R}$. But, by item \ref{enum:s2.res2}
in the definition of resonance, all lines external
to $W$ have the same scale $n^{R}_{W}$ except at most one.
\end{proof}

We define the \emph{resonance family} $\effe_{V}(\th)$ of $V\in\vres$ in
$\th$ as the set of trees obtained from $\th$ by the action of a group of
transformations $\pii_{V}$ on $\th$, generated by the following operations:
\begin{enumerate}
	\item\label{enum:s3.ren1} Detach the line $\ell_{1}$, then
	if $\ell_1\in L_{V}^{R}$ reattach it to all nodes internal to
	$V_0(\ell_{1})$, while if $\ell_{1}\in L_{V}^{0}$ reattach it
	to all nodes in $V_{0}$; for each 
	tree so obtained, do the same operations with the line 
	$\ell_{2}$ and so forth for each line entering the resonance.

	\item\label{enum:s3.ren2} In a given tree, each node $u \in V$ 
	will have $m_{u}$ entering lines, of which $s_{u}$ are inside $V$ 
	and $r_{u} = m_{u} - s_{u}$ are outside $V$ (\ie are entering 
	lines of $V$); then we can apply to the set of lines entering $u$ 
	a transformation in the group obtained as the quotient of the 
	group of permutations of the $m_{u}$ lines entering $u$ by the 
	groups of permutations of the $s_{u}$ internal entering lines and 
	of permutations of the $r_{u}$ entering lines outside $V$; in this 
	way for each node $u \in V$ a number of trees equal to:
\begin{equation*}
	\binom{m_{u}}{s_{u}} = \frac{m_{u}!}{s_{u}!r_{u}!}
\end{equation*}
	is obtained.

	\item\label{enum:s3.ren3} Flip simultaneously all the mode labels 
	of the nodes internal to $V$.
\end{enumerate}

We shall call \emph{renormalization transformations} (of type 
\ref{enum:s3.ren1}, \ref{enum:s3.ren2}, \ref{enum:s3.ren3})
the operations described above.

\begin{rem}\label{rem:s3.struttura}
	Note that in all such transformations the scales are not
	changed (by definition) and the set of resonance $\vres$
	remains the same (by construction). On the contrary
	the momenta flowing through the lines can change (because
	of the shift of the lines entering resonances) and in
	particular one can have for some lines $\ell$,
	$\chi_{n_{\ell}}(||\om\nu_{\ell}||) =0$, if
	$\nu_{\ell}$ is the modified momentum flowing through $\ell$.
\end{rem}

\begin{rem}
    The definition of resonance families is aimed at
    grouping together the trees between which one will look for
    compensations, but in doing so one has to avoid overcountings.
    In fact, to each tree $\th$ we associate a value 
    $\Val(\th)$ according to \eqref{eq:s2.treeexp}; when applying the 
    transformations of the group $\pii_{V}$ on the tree $\th$, the same 
    tree $\th'$ can be obtained, in general, in several ways; however, 
    it has to be counted once. This means that $\pii_{V}$, as a
    group, defines an equivalence class, and only inequivalent
    elements obtained through the transformations defining $\pii_{V}$
    have to be retained.
\end{rem}

Let us call $\effe_{\vres_{1}}(\th)$ the family obtained by
the composition of all transformations defining
the resonance families $\effe_{V_1}(\th)$, $V_1\in\vres_{1}$.

For any tree $\th_1\in\effe_{\vres_1}(\th)$, let $V_2$ be a resonance in
$\vres_{2}$ and let us define the resonance family
$\effe_{V_2}(\th_1)$ of $V_2$ in $\th_1$ as the set of trees obtained
from $\th_1$ by the action of the group of transformations
$\pii_{V_2}$. The composition of all transformations defining
the resonance families $\effe_{V_2}(\th_1)$, for all
$\th_1\in \effe_{\vres_{1}}(\th)$ and all $V_2\in\vres_{2}$,
gives a family that we shall denote by $\effe_{\vres_{2}}(\th)$.

We continue by considering resonances of $3$-rd generation,
and so on until the $G$-th generation resonances are reached.
At the end we shall have a family $\effe(\th)$ of trees
obtained by the composition of all transformations
of the groups $\pii_{V}$, $V\in \vres$, defined recursively
through the application of the renormalization transformations
corresponding to resonances $V\in\vres_{j}$ to all trees
$\th'$ belonging to the family $\effe_{\vres_{j-1}}(\th)$.

\begin{rem}\label{rem:s3.null}
	Given a tree $\th\in\ti_{\nu,k}$ and a family $\effe(\th)$,
	when considering another tree $\th'\in\effe(\th)$ with
	nonvanishing value $\Val(\th')$, the same family
	$\effe(\th')=\effe(\th)$ is obtained (by construction).
	Note however that $\effe(\th)$ can contain also trees
	with vanishing values, as they can have lines $\ell$
	such that $\chi_{n_{\ell}}(||\om \nu_{\ell}||)=0$
	(see remark \ref{rem:s3.struttura}).
\end{rem}

Define also $\enne_{\effe(\th)}$ the number of trees in $\effe(\th)$
whose value is not vanishing; of course $\enne_{\effe(\th)}\leq
|\effe(\th)|$, if $|\effe(\th)|$ is the number of elements in $\effe(\th)$.

Write:
\begin{equation}
		\sum_{\th\in\ti_{\nu,k}} \Val(\th) =
		\sum_{\th\in\ti_{\nu,k}} \frac{1}{\enne_{\effe(\th)}}
		\sum_{\th'\in\effe(\th)} \Val(\th') =
		\sum_{\th\in\ti_{\nu,k}^*} \frac{1}{|\effe(\th)|}
		\sum_{\th'\in\effe(\th)} \Val(\th') ,
	\label{eq:s4.eq0}
\end{equation}
where the factors $\enne_{\effe(\th)}$ and $|\effe(\th)|$ have been
intoduced in order to avoid overcountings (see remark \ref{rem:s3.null})
and the last sum implicitly defines the set $\ti_{\nu,k}^*$:
so $\ti_{\nu,k}^*$ is the set of inequivalent trees in
$\cup_{\th\in\ti_{\nu,k}}\effe(\th)$.

Consider a tree $\th\in\ti_{\nu,k}^*$. Then $\th\in \effe(\th_0)$, for
some tree $\th_0\in\ti_{\nu,k}$; however one has to bear in mind that
$\th$, unlike $\th_0$, could vanish.

\vspace{.5truecm}

Given a tree $\th\in\ti_{\nu,k}^*$, if $V$ is a
first generation resonance, we define its \emph{resonance 
factor} $\vu_{V}(\th)$ as its contribution to the value of the tree 
$\th$, namely:
\begin{equation}
	\vu_{V}(\th) = \Biggl[
		\prod_{u \in V}\frac{\nu_{u}^{m_{u}+1}}{m_{u}!}
		\Biggr] \Biggl[
		\prod_{\ell \in V}g_{n_{\ell}}(\nu_{\ell})
		\Biggr],
	\label{eq:s3.resfact}
\end{equation}
which of course depends on the subset of $\th$ outside the resonance 
$V$ only through the momenta of the entering lines of $V$. Given a 
node $u \in V$, let us denote with $\ee_{u}$ the set of lines 
entering $V$ such that they end into nodes preceding $u$. 

For future notational convenience, we rewrite \eqref{eq:s3.resfact} as:
\begin{equation}
	\vu_{V}(\th) = U_{V}(\th)L_{V}(\th), \quad 
	U_{V}(\th) = \prod_{u \in V}\frac{\nu_{u}^{m_{u}+1}}{m_{u}!}, \quad 
	L_{V}(\th) = \prod_{\ell \in V}g_{n_{\ell}}(\nu_{\ell}).
	\label{eq:s3.resfact2}
\end{equation}

In the following, we shall consider the quantities $\om\nu$, $\nu \in 
\fZ$, modulo $1$, and shall continue to use the symbol $\om\nu$ to 
denote the representative of the equivalence class within the 
interval $(-1/2,1/2]$. 

For any node $u$ contained in a resonance $V$, we shall write:
\begin{equation}
    \nu_{\ell_{u}} = \nu^{0}_{\ell_{u}} + \sum_{\ell' \in 
    	\ee_{u}}\nu_{\ell'}, \quad 
    \nu^{0}_{\ell_{u}} = \sum_{\substack{w \in V\\ w \tle u}}\nu_{w},
    \label{eq:s3.nuzero}
\end{equation}
where the set $\ee_{u}$ was defined after \eqref{eq:s3.resfact}. 

We shall consider the resonance factor \eqref{eq:s3.resfact} as a 
function of the quantities $\mu_{1} = \om\nu_{\ell_{1}}$, \dots, 
$\mu_{m_{V}} = \om\nu_{\ell_{m_{V}}}$, where $\nu_{\ell_{1}}$, \dots, 
$\nu_{\ell_{m_{V}}}$ are the momenta flowing through the lines 
$\ell_{1}$, \dots, $\ell_{m_{V}}$ entering $V$. More precisely, we let:
\begin{equation}
    \vu(\th) \equiv
\vu_{V}(\th;\om\nu_{\ell_{1}},\dots,\om\nu_{\ell_{m_{V}}}),
    \label{eq:s3.notation}
\end{equation}
and we write:
\begin{equation}
	\begin{split}
	& \vu_{V}(\th;\om\nu_{\ell_{1}},\dots,\om\nu_{\ell_{m_{V}}}) = \\
	& \qquad =
	\elle\vu_{V}(\th;\om\nu_{\ell_{1}},\dots,\om\nu_{m_{V}}) + 
	\erre\vu_{V}(\th;\om\nu_{\ell_{1}},\dots,\om\nu_{m_{V}}),
	\label{eq:s3.locren}
	\end{split}
\end{equation}
where:
\begin{equation}
	\begin{split}
	& \elle\vu_{V}(\th;\om\nu_{\ell_{1}},\dots,\om\nu_{\ell_{m_{V}}}) = \\
	& \qquad = \vu_{V}(\th;0,\dots,0) + 
	\sum_{m=1}^{m_{V}}\om\nu_{\ell_{m}}\frac{\p}{\p\mu_{m}}
	\vu_{V}(\th;0,\dots,0)
	\label{eq:s3.locdef}
	\end{split}
\end{equation}
is the \emph{localized part} of the resonance factor, or 
\emph{localized resonance factor}, while:
\begin{multline}
	\erre\vu_{V}(\th;\om\nu_{\ell_{1}},\dots,\om\nu_{\ell_{m_{V}}}) = 
	\sum_{m,m'=1}^{m_{V}}\om\nu_{\ell_{m}}\,\om\nu_{\ell_{m'}} 
	\cdot \\ \cdot
	\int_{0}^{1}\der t\; (1-t)\frac{\p^{2}}{\p\mu_{m}\p\mu_{m'}}
	\vu_{V}(\th;t\om\nu_{\ell_{1}},\dots,t\om\nu_{\ell_{m_{V}}})
	\label{eq:s3.rendef}
\end{multline}
is the \emph{renormalized part} of the resonance factor, or 
\emph{renormalized resonance factor}. In \eqref{eq:s3.locren} $\elle$ 
is called the \emph{localization operator} and $\erre = 1 - \elle$ is 
called the \emph{renormalization operator}. Using the notations 
\eqref{eq:s3.resfact2}, we can write:
\begin{equation}
	\elle\vu_{V}(\th) = U_{V}(\th)\elle L_{V}(\th), \quad 
	\erre\vu_{V}(\th) = U_{V}(\th)\erre L_{V}(\th),
	\label{eq:s3.sololelinee}
\end{equation}
as only the factors in $L_{V}(\th)$ depend on the momenta flowing 
through the lines entering the resonance $V$. 

\begin{rem}\label{rem:s3.locpartvan}
Note that in the localized part \eqref{eq:s3.locdef} the momentum 
$\nu_{\ell}$ flowing through any line $\ell$ internal to $V$ is 
changed into $\nu^{0}_{\ell}$ (see \eqref{eq:s3.nuzero}).
\end{rem}

Then we perform the renormalization transformations
in $\pii_{V}$ described above. By remark \ref{rem:s3.locpartvan},
for all trees obtained by applying the group 
$\pii_{V}$ the contribution to the localized resonance factor arising 
from the $L_{V}(\th)$ term in \eqref{eq:s3.resfact2} is the same, \ie:
\begin{equation}
	\elle L_{V}(\th) = \elle L_{V}(\th'), \quad 
	\forall \th' \in \effe_{V}(\th),
	\label{eq:s3.caso2-1}
\end{equation}
so that we can consider:
\begin{equation}
	\sum_{\th' \in \effe_{V}(\th)} \elle \vu_{V}(\th').
	\label{eq:s3.resum}
\end{equation}
The sum over all the trees in the resonance family $\effe_{V}(\th)$
of the localized resonance factors produces zero,
so that only the renormalized part has to be taken into 
account. The proof of this assertion is similar to the proof of the 
analogous statement in \cite{BG1}, and it is given
in sect. \ref{sect:proof6} as a particular case of the 
proof of the more general statement in lemma \ref{lem:s4.lemma6} below. 

Then only the second order terms have to be taken into account in 
\eqref{eq:s3.locren}. This leads to the following expression
for the renormalized resonance factor:
\begin{multline}
		\erre\vu_{V}(\th) = U_V(\th)\sum_{m,m' = 1}^{m_{V}}
		\om\nu_{\ell_{m}}\,\om\nu_{\ell_{m'}}\cdot\\
		\quad \cdot \Biggl[
		\sum_{\substack{\ell^{1}_{V},\ell^{2}_{V}\in V\\
		\ell^{1}_{V}\neq\ell^{2}_{V}}}
		\biggl(\frac{\p}{\p\mu_{m}}g_{n_{\ell^{1}_{V}}}
		(\nu_{\ell^{1}_{V}})\biggr)
		\biggl(\frac{\p}{\p\mu_{m'}}g_{n_{\ell^{2}_{V}}}
		(\nu_{\ell^{2}_{V}})\biggr)
		\biggl(\prod_{\substack{\ell\in V\\
		\ell\neq\ell^{1}_{V},\ell^{2}_{V}}}
		g_{n_{\ell}}(\nu_{\ell})\biggr) + \\
		\quad + \sum_{\ell_{V}\in V}
		\biggl(\frac{\p}{\p\mu_{m}}\frac{\p}{\p\mu_{m'}}
		g_{n_{\ell_{V}}}(\nu_{\ell_{V}})\biggr)
		\biggl(\prod_{\substack{\ell\in V\\
		\ell\neq\ell_{V}}}g_{n_{\ell}}(\nu_{\ell})\biggr)\Biggr],
	\label{eq:s3.lem5}
\end{multline}
from the very definition of the renormalized resonance factor 
\eqref{eq:s3.rendef}, by noting that the two derivatives in
\eqref{eq:s3.rendef} act either on two distinct propagators (the sum with
$\ell^{1}_{V}\neq\ell^{2}_{V}$ in \eqref{eq:s3.lem5}) or on the same
propagator (the sum with only one line $\ell_{V}$ in \eqref{eq:s3.lem5}).

Note that it can happen that $\th\in\effe_V(\th_0)$,
for some tree $\th_0\in\ti_{\nu,k}$, \emph{i.e.} for
some tree $\th_0$ with nonvanishing value, while $\vu_{V}(\th)=0$
(correspondingly there does not exist any tree in $\ti_{\nu,k}$
of that shape associated with the given choice of mode and scale labels).
The tree $\th$ is obtained from $\th_0$ through a transformation
of $\pii_{V}$, so that there is a
correspondence between the lines of $\th_0$ and the lines of $\th$:
we shall say that the lines are \emph{conjugate}.
The tree $\th$ inherits the scale
labels of the tree $\th_0$, \emph{i.e} the lines in $\th$
have the same scales of the conjugate lines of $\th_0$.
So it can happen that in $\th_0$
some line internal to $V$ has a scale $n_{\ell}$ and a
momentum $\tilde\nu_{\ell}$ such that
$\chi_{n_\ell}(||\om\tilde\nu_{\ell}||)\neq 0$,
while the momentum $\nu_{\ell}$ of the line $\ell$ seen as a line
of $\th$ (\emph{i.e.} of the line of $\th$
conjugate to the line $\ell$ of $\th_0$)
is such that $\chi_{n_{\ell}}(||\om\nu_{\ell}||)=0$
(see remark \ref{rem:s3.null}).
This means that for such a line \eqref{eq:s2.distest} does not hold.
Nevertheless, as anticipated in remark \ref{rem:s3.struttura},
one finds that the momentum $\nu_{\ell}$ can not change
``too much'' with respect to $\tilde\nu_{\ell}$; more precisely:
\begin{equation}
		\frac{1}{768q_{n_{\ell}+1}} \leq
			||\om\nu_{\ell}|| \leq 
			\frac{1}{24q_{n_{\ell}}},
		\label{eq:s3.distest11}
\end{equation}
as we shall prove, using the following result.

\begin{lem}\label{lem:s3.cambio}
	Given a tree $\th_0\in\ti_{\nu,k}$ and a resonance $V$,	let
	$\th\in\ti_{\nu,k}^*$ be a tree obtained by the action of the
	group $\pii_{V}$, \emph{i.e.} $\th\in\effe_{V}(\th_0)$.
	If $||\om\nu_{\ell_{m}}||\leq 1/8q_{n_{V}^{R}}$
	for any entering line $\ell_{m}$ of $V$,
	$m=1,\ldots,m_{V}$, then, for any line $\ell\in V$, one has
	\begin{equation}
		\bigl| ||\om\nu_{\ell}|| - ||\om\tilde\nu_{\ell}|| \bigr|
		\leq \frac{1}{4q_{n_{V}^{R}}} , \qquad ||\om\nu_{\ell}||\geq
		\frac{1}{4q_{n_{V}^{R}}} , \qquad ||\om\tilde\nu_{\ell}||\geq 
		\frac{1}{4q_{n_{V}^{R}}} ,
		\label{eq:s3.f}
	\end{equation}
	if $\nu_{\ell}$ and $\tilde\nu_{\ell}$ are the momenta flowing
	through $\ell$ in $\th$ and $\th_0$, respectively.
\end{lem}

\begin{proof}
As $V$ is a resonance, then for each line $\ell\in V$ one
has $|\nu_{\ell}^0|\leq k_V< q_{n_{V}^{R}}$ (see item \ref{enum:s2.res4}
in the definition of resonance), so that
\begin{equation}
	||\om\nu_{\ell}^0|| \geq ||\om q_{n_{V}^{R}-1}||
	> \frac{1}{2q_{n_{V}^{R}}} ,
	\label{eq:s3.a}
\end{equation}
by \eqref{eq:s2.tmp1} and \eqref{eq:s2.tmp2}.
On the other hand
\begin{equation}
	|| \om\nu_{\ell}-\om\nu_{\ell}^0 || \leq
	\sum_{m=1}^{m_V}||\om\nu_{m} || , 
	\label{eq:s3.b}
\end{equation}
if $\nu_1,\ldots,\nu_{m_V}$ are the momenta flowing through the
lines $\ell_1,\ldots,\ell_{m_V}$ entering $V$. By hypothesis
\begin{equation}
	|| \om\nu_{\ell_m}|| \leq \frac{1}{8q_{n_{V}^{R}}} ,
	\quad \forall m=1,\ldots,m_V .
	\label{eq:s3.c}
\end{equation}
If $m_{V}\ge 2$ then one must have $q_{n_{V}^{R}+1}>4q_{n_{V}^{R}}$
(see item \ref{enum:s2.res5} in the definition of resonance).
In such a case if there is an entering line
(say $\ell_{1}$) which is the root line of a
tree of order $\geq q_{n_{V}^{R}+1}/4$, then all the other lines are
the root lines of subtrees of orders $k_2,\ldots,k_{m_V}$ such that
$k_0\equiv k_2+\ldots+k_{m_V}<q_{n_{V}^{R}+1}/8$
(see item \ref{enum:s2.res61} in the definition of resonance).
Moreover, for each $m=2,\ldots,m_{V}$,
$k_m\geq q_{n_{V}^{R}}$, otherwise the line $\ell_m$ would not be on scale
$\geq n_{V}^{R}$. By lemma \ref{lem:s2.davie}, $\nu_{m}=s_{m}q_{n_{V}^{R}}$
for all $m=2,\ldots,m_V$, with $s_m\in\fZ$, and
\begin{equation}
	|s_2|+\ldots+|s_{m_V}| \leq \frac{k_0}{q_{n_{V}^{R}}}
	\leq \frac{q_{n_{V}^{R}+1}}{8q_{n_{V}^{R}}},
	\label{eq:s3.d}
\end{equation}
so that
\begin{equation}
	\sum_{m=1}^{m_V} ||\om\nu_m|| \leq \frac{1}{8q_{n_{V}^{R}}}+
	\sum_{m=2}^{m_V}|s_m| \, ||\om q_{n_{V}^{R}}|| \leq
	\frac{1}{8q_{n_{V}^{R}}}+ \frac{1}{8q_{n_{V}^{R}}}
	\leq \frac{1}{4q_{n_{V}^{R}}} ,
	\label{eq:s3.e}
\end{equation}
where use was made of \eqref{eq:s2.tmp1}.
Therefore, when replacing $\th_0$ with $\th$,
\eqref{eq:s3.f} follows.

If there is no entering line of $V$ which is the root line of a tree
of order $\geq q_{n_{V}^{R}+1}/4$ and the tree having as root line the
exiting line of $V$ is of order $k< q_{n_{V}^{R}+1}/4$
(see item \ref{enum:s2.res62} in the definition of resonance), then
\begin{equation}
	\sum_{m=1}^{m_{V}} |s_m|q_{n_{V}^{R}} \leq k_1+\ldots+k_{m_V}
	\equiv k -k_T < k \leq \frac{q_{n_{V}^{R}+1}}{4} \; , 
		\label{eq:s3.g}
\end{equation}
so that
\begin{equation}
	\sum_{m=1}^{m_V} ||\om\nu_m|| \leq
	\sum_{m=1}^{m_V}|s_m| \, ||\om q_{n_{V}^{R}}|| \leq
	\frac{q_{n_{V}^{R}+1}}{4q_{n_{V}^{R}}} \frac{1}{q_{n_{V}^{R}+1}}
	=\frac{1}{4q_{n_{V}^{R}}} .
	\label{eq:s3.h}
\end{equation}
which implies again \eqref{eq:s3.f}.
If $m_V=1$, then \eqref{eq:s3.f} follows immediately from
\eqref{eq:s3.b} and \eqref{eq:s3.c}.
\end{proof}


We come back to the proof of \eqref{eq:s3.distest11}. Note that inside
$V$ in $\th_0$ (hence also in $\th$, see remark \ref{rem:s3.struttura})
only lines on scale $n_{\ell}$ such that
$1/48q_{n_{\ell}}>1/4q_{n_{V}^{R}}$ are possible, by the second
inequality in \eqref{eq:s3.f} and the definition of scale
(see \eqref{eq:s2.distest}).

As the entering lines of $V$ satisfy \eqref{eq:s2.distest},
hence \eqref{eq:s2.distest1elle}, lemma \ref{lem:s3.cambio} applies.
Then, given a line $\ell$ internal to $V$ on scale $n_{\ell}$, one has
\begin{equation}
	||\om\nu_{\ell}||\leq \frac{1}{48q_{n_{\ell}}}
	+ \frac{1}{4q_{n_{V}^{R}}} \leq \frac{1}{48q_{n_{\ell}}} +
	\frac{1}{48q_{n_{\ell}}} \leq \frac{1}{24q_{n_{\ell}}} .
	\label{eq:s3.j}
\end{equation}

Likewise, if $1/96q_{n_{\ell}+1}>2/q_{n_{V}^{R}}$, one has
\begin{equation}
	||\om\nu_{\ell}||\geq \frac{1}{96q_{n_{\ell}+1}}
	- \frac{1}{4q_{n_{V}^{R}}} \geq \frac{1}{96q_{n_{\ell}+1}} -
	\frac{1}{768q_{n_{\ell}+1}} \geq \frac{1}{96q_{n_{\ell}+1}}
	\Biggl( 1-\frac{1}{8} \Biggr) ,
	\label{eq:s3.k}
\end{equation}
while, if $1/96q_{n_{\ell}+1}<2/q_{n_{V}^{R}}$, one has
\begin{equation}
	||\om\nu_{\ell}||\geq \frac{1}{4q_{n_{V}^{R}}}
	\geq \frac{1}{768q_{n_{\ell}+1}} \; .
	\label{eq:s3.l}
\end{equation}
by the third inequality in \eqref{eq:s3.f}.
Then \eqref{eq:s3.distest11} follows: so in particular the
momentum $\nu_{\ell}$ of the line $\ell\in\th$
still fulfills \eqref{eq:s2.distest1elle}.

\vspace{.5truecm}

Note that \eqref{eq:s3.lem5} and \eqref{eq:s2.distest1elle} imply
the following bound for the renormalized resonance factor
of a first generation resonance:
\begin{equation}
	\begin{split}
		|\erre\vu_{V}(\th)| \leq D_{6}D_{7}^{k_{V}} &
		\sum_{m,m'=1}^{m_{V}}||\om\nu_{\ell_{m}}||\,
		||\om\nu_{\ell_{m'}}|| \cdot \\
		& \cdot \bigl(768q_{n_{V}+1}\bigr)^{2}
		\biggl(\prod_{\ell\in V}
		\bigl(768q_{n_{\ell}+1}\bigr)^{2}\biggr),
	\end{split}
	\label{eq:s3.renbound}
\end{equation}
(for some constants $D_{6}$ and $D_{7}$),
where the last product (times $\Gamma^{-k}$) represents a bound on the 
resonance factor \eqref{eq:s3.resfact}. The proof of such an 
assertion again is as in \cite{BG1} (see the proof of the Corollary 
in \cite{BG1}, \S 3), and follows immediately by noting that for
any line $\ell \in V$ one has $n_{\ell} \geq n_{V}$.
The only difference with respect to \cite{BG1} 
is that now the derivatives can act also on the compact support 
functions: they were just missing in \cite{BG1}; it is nevertheless 
straightforward to see that:
\begin{equation}
	\Biggl|
	\frac{\p^{p}}{\p^{p}\mu}\chi_{n}(||\om\nu_{\ell}||)
	\Biggr| \leq D_{8}\bigl(768q_{n+1}\bigr)^{p},
	\label{eq:s3.charfunest1}
\end{equation}
with $p=1,2$, for some constant $D_{8}$, so that:
\begin{equation}
	\Biggl|
	\frac{\p^{p}}{\p^{p}\mu}g_{n}(\nu_{\ell})
	\Biggr| \leq D_{9}\bigl(768q_{n+1}\bigr)^{p+2},
	\label{eq:s3.charfunest2}
\end{equation}
with $p=0,1,2$, for some constant $D_{9}$. 

For any tree in $\effe_{V}(\th)$ the bound \eqref{eq:s2.distest1elle}
holds, so that lemma \ref{lem:s2.count} applies (see remark
\ref{rem:s5.rem14} in sect. \ref{sect:proof4}).

Note that the two factors $||\om\nu_{\ell_{m}}||$, 
$||\om\nu_{\ell_{m'}}||$ in \eqref{eq:s3.renbound} allow us to 
neglect the propagator corresponding to a line entering a resonance 
with resonance scale $n^{R}_{V}$, provided such a propagator is 
replaced by a factor $(768q_{n_{V}+1})^{2}$, where $n_{V}$ is the 
scale of the resonance as a cluster. Such a mechanism corresponds 
to the discussion leading to \eqref{eq:s2.trivbound3}, as far as only 
the first generation resonances are considered. 

In general a tree will contain more resonances, and the resonances 
can be contained into each other. Then the above
discussion has to be extended to cover the more general case:
which will be done in the next section.

\vspace{1cm}

\section{Renormalization of resonances: the general step}
\label{sect:canc3}

\noindent
We proceed following strictly the techniques of \cite{GM} and \cite{BGGM}.

Consider a tree $\th\in\ti_{\nu,k}^*$ in \eqref{eq:s4.eq0}.
For each resonance $V$ of any generation, let us define a pair of
\emph{derived lines} $\ell^{1}_{V}$, $\ell^{2}_{V}$ internal to $V$
-- possibly coinciding -- with the following ``compatibility'' condition:
if $V$ is inside some other resonance $W$, the set 
$\{\ell^{1}_{V},\ell^{2}_{V}\}$ must contain those lines of 
$\{\ell^{1}_{W},\ell^{2}_{W}\}$ which are inside $V$.  Clearly there 
can be $0$, $1$ or $2$ such lines, and correspondingly we shall say 
that the resonance $V$ is of type $2$ if none of its derived lines is 
a derived line for one of the resonances containing it, of type $1$ if
just one of its two derived lines is a derived line for one of the 
resonances containing it, and of type $0$ if both derived lines are 
derived lines for some resonances $W$, $W'$ -- possibly coinciding -- 
containing $V$; we shall use a label $z_{V} = 0, 1, 2$ to take note of 
the type of the resonance $V$.  One associates also to each resonance 
$V$ a pair of entering lines $\ell_{m}^{V}$, $\ell_{m'}^{V}$ if 
$z_{V}=2$ and a single line $\ell_{m}^{V}$ if $z_{V}=1$, with $m, m' = 
1$, \dots, $m_{V}$. Moreover for each resonance
we shall introduce an interpolation parameter $t_{V}$ and a measure 
$\pi_{z_V}(t_V)\,\der t_{V}$ such that:
\begin{equation}
	\pi_{z}(t) = \begin{cases}
    (1-t), &\quad z=2\\
    1, &\quad z=1\\
    \delta(t-1), &\quad z=0;
    \end{cases}
	\label{eq:s4.meas}
\end{equation}
we shall denote with $\mathbf{t} = \{t_{V}\}_{V\in\vres}$
the set of all interpolation parameters.

The momentum flowing through a line $\ell_{u}$ internal to any resonance $V
$
will be defined recursively as:
\begin{equation}
	\nu_{\ell_{u}}(\mathbf{t})  =  \nu^{0}_{\ell_{u}} +
	t_{V}\sum_{\ell\in\ee_{u}}\nu_{\ell}(\mathbf{t}), \quad
	\nu^{0}_{\ell_{u}} = \sum_{\substack{w\in V\\ w \tle u}}
	\nu_{w};
	\label{eq:s4.defnuellu}
\end{equation}
of course $\nu_{\ell_{u}}(\mathbf{t})$ will depend only on the 
interpolation parameters corresponding to the resonances containing 
the line $\ell_{u}$ (by construction). 

For any resonance $V$ the resonance factor is defined as
\begin{equation}
	\vu_{V}(\th) = U_{V}(\th)\Biggl[
		\prod_{\ell\in V}g_{n_{\ell}}(\nu_{\ell}(\mathbf{t}))
		\Biggr] ,
	\label{eq:s4.lv11}
\end{equation}
when $z_V=2$, as
\begin{equation}
	\vu_{V}(\th) = U_{V}(\th)\Biggl[
		\Bigl(\frac{\p}{\p\mu}
		g_{n_{\ell^{1}_{V}}}(\nu_{\ell^{1}_{V}}(\mathbf{t}))\Bigr)
		\biggl(\prod_{\substack{\ell\in V,\\\ell\neq\ell^{1}_{V}}}
		g_{n_{\ell}}(\nu_{\ell}(\mathbf{t}))\biggr)\Biggr] ,
	\label{eq:s4.lv12}
\end{equation}
when $z_V=1$ (and we have called $\ell^1_V$ the line
in $\{\ell_V^1,\ell_V^2\}$ which belongs
to the set $\{\ell_W^1,\ell_W^2\}$ for some
resonance $W$ containing $V$), as
\begin{equation}
	\vu_{V}(\th) = U_{V}(\th)\Biggl[
		\Bigl(\frac{\p^{2}}{\p\mu\p\mu'}
		g_{n_{\ell^{1}_{V}}}(\nu_{\ell^{1}_{V}}(\mathbf{t}))\Bigr)
		\biggl(\prod_{\substack{\ell\in V,\\\ell\neq\ell^{1}_{V}}}
		g_{n_{\ell}}(\nu_{\ell}(\mathbf{t}))\biggr)\Biggr] ,
	\label{eq:s4.lv13-1}
\end{equation}
when $z_V=0$ and $\ell^{1}_{V} = \ell^{2}_{V}$, and as
\begin{multline}
	\vu_{V}(\th) = U_{V}(\th)\Biggl[
		\Bigl(\frac{\p}{\p\mu}
		g_{n_{\ell^{1}_{V}}}(\nu_{\ell^{1}_{V}}(\mathbf{t}))\Bigr)
		\Bigl(\frac{\p}{\p\mu'}
		g_{n_{\ell^{2}_{V}}}(\nu_{\ell^{2}_{V}}(\mathbf{t}))\Bigr)
		\cdot\\
		\cdot\biggl(\prod_{\substack{\ell\in
V,\\\ell\neq\ell^{1}_{V},\ell^{2}_{V}}}
		g_{n_{\ell}}(\nu_{\ell}(\mathbf{t}))\biggr)\Biggr] ,
	\label{eq:s4.lv13-2}
\end{multline}
when $z_V=0$ and $\ell^{1}_{V} \neq \ell^{2}_{V}$.

In \eqref{eq:s4.lv12}$\div$\eqref{eq:s4.lv13-2} one has
$\mu=\om\nu_{\ell_{m}^{W}}$ and $\mu'=\om\nu_{\ell_{m'}^{W'}}$,
for some lines $\ell_{m}^{W}$ and $\ell_{m'}^{W'}$ (possibly coinciding)
entering, respectively, some resonances $W$ and $W'$
(possibly coinciding) containing $V$.

We define the renormalization operator according to the type of the 
resonance; namely, if $z_{V} = 2$, then:
\begin{multline}
	\erre\vu_{V}(\th;\om\nu_{\ell_{1}}(\mathbf{t}),\dots,
		\om\nu_{\ell_{m_{V}}}(\mathbf{t})) = \sum_{m,m'=1}^{m_{V}}
		\om\nu_{\ell_{m}}(\mathbf{t})\om\nu_{\ell_{m'}}
		(\mathbf{t})\cdot \\
		\cdot\int_{0}^{1}\der t_{V}\,(1-t_{V})
		\frac{\p^{2}}{\p\mu_{m}\p\mu_{m'}}
		\vu_{V}(\th,t_{V}\om\nu_{\ell_{1}}(\mathbf{t}),\dots,
		t_{V}\om\nu_{\ell_{m_{V}}}(\mathbf{t}));
	\label{eq:s4.grendef1}
\end{multline}
if $z_{V} = 1$, then:
\begin{multline}
	\erre\vu_{V}(\th;\om\nu_{\ell_{1}}(\mathbf{t}),\dots,
		\om\nu_{\ell_{m_{V}}}(\mathbf{t})) = 
		\sum_{m=1}^{m_{V}}\om\nu_{\ell_{m}}(\mathbf{t})\cdot\\
		\cdot\int_{0}^{1}\der t_{V}\;\frac{\p}{\p\mu_{m}}
		\vu_{V}(\th,t_{V}\om\nu_{\ell_{1}}(\mathbf{t}),\dots,
		t_{V}\om\nu_{\ell_{m_{V}}}(\mathbf{t}));
	\label{eq:s4.grendef2}
\end{multline}
finally if $z_{V} = 0$, then:
\begin{equation}
	\erre\vu_{V}(\th)(\th;\om\nu_{\ell_{1}}(\mathbf{t}),\dots,
		\om\nu_{\ell_{m_{V}}}(\mathbf{t})) = 
		\vu_{V}(\th)(\th;\om\nu_{\ell_{1}}(\mathbf{t}),\dots,
		\om\nu_{\ell_{m_{V}}}(\mathbf{t})).
	\label{eq:s4.grendef3}
\end{equation}
In all cases set $\elle = 1 - \erre$. 

\begin{rem}\label{rem:s4.rem2}
Note that $z_{V}$ equals the order of the renormalization performed on 
the resonance $V$. 
\end{rem}

\begin{rem}
	If a resonance $V$ has a resonance scale $n^{R}_{V}$, then there is 
	a line $\ell^{0}_{V}$ on scale $n^{R}_{V}$ entering $V$ such that 
	$||\om\nu_{\ell}|| \leq ||\om\nu_{\ell^{0}_{V}}||$ for each $\ell$ 
	entering $V$. If there is ambiguity, $\ell^{0}_{V}$ can be chosen 
	arbitrarily.
	For any resonance $V$ one has a factor bounded by 
	$||\om\nu_{\ell^{0}_{V}}||^{z_{V}}$, from \eqref{eq:s4.grendef1}, 
	\eqref{eq:s4.grendef2} and \eqref{eq:s4.grendef3} and by the 
	definition of $\ell^{0}_{V}$. 
\end{rem}

To each line $\ell$ derived once one can associate the line 
$\ell_{m}(\ell)$ corresponding to the quantity $\mu_{m} = 
\om\nu_{\ell_{m}(\ell)}$ with respect to which the propagator 
$g_{n_{\ell}}(\nu_{\ell}(\mathbf{t}))$ is derived.
If the line $\ell$ is derived twice one associates to it
the two lines $\ell_{m}(\ell)$ and $\ell_{m'}(\ell)$
such that $\mu_{m}=\om\nu_{\ell_{m}(\ell)}$ and
$\mu_{m'} = \om\nu_{\ell_{m'}(\ell)}$ are the quantities
with respect to which the propagator
$g_{n_{\ell}}(\nu_{\ell}(\mathbf{t}))$ is derived.

Given a	derived	line $\ell$, let $V$ be	the	minimal	resonance 
containing it. If the line $\ell$ is derived once, then let $W$ be the 
resonance for which $\ell_{m}(\ell)$ is an entering line; if instead 
$\ell$ is derived twice, let $W, W' \subseteq W$ be the resonances for 
which the lines $\ell_{m}(\ell)$, $\ell_{m'}(\ell)$ respectively are 
entering lines. 

In the first case, let $W_{i}$, $i=0$, \dots, $p$ the resonances contained
by $W$ and containing $V$, ordered naturally by	inclusion:
\begin{equation}															
	V =	W_{0} \subset W_{1}	\subset	\dots \subset W_{p}	= W.				
	\label{eq:s4.cloud1}													
\end{equation}																
We shall call the set $\wres(\ell) = \{W_{0},\dots,W_{p}\}>$
the \emph{simple cloud} of $\ell$.	

In the second case, let $W_{i}$, $i=0$, \dots, $p$, the resonances 
contained by $W$ and containing $V$, ordered naturally by inclusion:
\begin{equation}
	V =	W_{0} \subset W_{1}	\subset	\dots \subset W_{p'} = W'
	\subset \dots \subset W_{p}	= W,
	\label{eq:s4.cloud2}
\end{equation}
with $p' \leq p$.  We shall say that $\wres_{-}(\ell) = 
\{W_{0},\dots,W_{p'}\}$ is the \emph{minor cloud} of $\ell$ while 
$\wres_{+}(\ell)=\{W_{0},\dots,W_{p}\}$ is the \emph{major cloud} of $V$.

When the renormalization of a resonance $V\in \vres_{j+1}$
is performed, a tree $\th^V_0\in\effe_{V_{j}}(\th)$,
with $\th\in\ti_{\nu,k}$, is replaced by the action of the group
$\pii_{V}$ with a new tree $\th^V$. As this replacement
is performed iteratively, one has the constraint that
if $V_1$ and $V_2$ are two resonance such that
$V_1$ is the minimal resonance containing $V_2$,
then $\th^{V_{1}}=\th_0^{V_{2}}$. At the end, the
original tree $\th_0\in\ti_{\nu,k}$
is replaced with a tree $\th\in\ti_{\nu,k}^*$. On each resonance
$V\in\vres$ of $\th$ the renormalization operator $\erre$ acts:
a tree whose resonance factors have been all renormalized will
be called a \emph{renormalized} (or \emph{resummed}) \emph{tree}.

As the replacement corresponding to each
resonance settles a conjugation between lines of $\th^{V}_{0}$ and
those of $\th^{V}$, in the end for each line of $\th$
there will be a conjugate line of $\th_{0}$.

Note that, as the transformations of the groups $\pii_{V}$,
$V\in\vres$, do not modify the scales of $\th_0$
(see remark \ref{rem:s3.struttura}),
the scales of the lines of $\th$ are the same as those of
the conjugate lines of the tree $\th_0$, so that, in order to apply
lemma \ref{lem:s2.count}, we have only to verify that
\eqref{eq:s2.distest1elle} is verified for the lines in $\th$:
this will be done below (after remark \ref{rem:s4.rem3}).

\vspace{.5truecm}

Now, we shall show that:

\begin{itemize}
	\item   the localized resonance factors can be
		neglected (in a sense that will appear clear shortly,
		see lemma \ref{lem:s4.lemma6} below),

	\item   for any (renormalized) resonance we obtain a factor:
		\begin{equation}
			\bigl(768q_{n_{V}+1}\bigr)^{2}
			||\om\nu_{\ell^{0}_{V}}||^{2},
			\label{eq:s4.renresbound}
		\end{equation}
	and

	\item   the number of terms generated by the renormalization
		procedure is bound\-ed by a costant to the power $k$,
\end{itemize}

\noindent so that the bound \eqref{eq:s2.trivbound2}
can be replaced by a bound which leads to \eqref{eq:s2.trivbound3},
as anticipated in sect. \ref{sect:trees}.

Note firstly that the localized part of the resonance factors can be 
dealt with as in sect.  \ref{sect:canc1}, when only first generation 
resonances were considered.  More formally, we have the following 
result, which is proved in sect. \ref{sect:proof6}.

\begin{lem}\label{lem:s4.lemma6}
Given a tree $\th$ and a resonance $V \in \th$, the localized 
resonance factor $\elle\vu_{V}(\th)$ gives zero 
when the values of the trees belonging to the same resonance family 
$\effe_{V}(\th)$ are summed together.
\end{lem}

Define the map $\Lambda$:
\begin{equation}
	\Lambda\!: \vres \mapsto \Lambda \vres =
	\bigl\{z_{V},\ell^{1}_{V},\ell_{V}^{2},
	\{\ell^{V}_{m},\ell^{V}_{m'}\}^{*}\bigr\}_{V\in\vres},
	\label{eq:s4.lambda}
\end{equation}
which associates to each resonance $V\in\vres$ the derived lines 
$\ell_{V}^{1},\ell_{V}^{2}$ and the lines in the set
$\{\ell^{V}_{m},\ell^{V}_{m'}\}^{*}$ defined as:
\begin{equation}
	\{\ell^{V}_{m},\ell^{V}_{m'}\}^{*} = \begin{cases}
		\{\ell^{V}_{m},\ell^{V}_{m'}\} , & \quad \text{if $z_{V}=2$},\\
		\ell^{V}_{m} , & \quad \text{if $z_{V}=1$},\\
		\emptyset , & \quad \text{if $z_{V}=0$},
	\end{cases} 
	\label{eq:s4.linee}
\end{equation}
where $m,m'=1$, \dots, $m_{V}$ and $\ell^{V}_{1}$, \dots, 
$\ell^{V}_{m_{V}}$ are the lines entering $V$. 

Note that that the map $\Lambda$ gives a natural decomposition of 
the set $L$ of all lines of $\th$ into $L=L_{0}\cup L_{1} \cup 
L_{2}$, where $L_{j}$ is the set of lines derived $j$ times.

Then, by using also lemma 6, one has
\begin{equation}
	\begin{split}
		\Val(\th) & =
		\sum_{\Lambda\vres} \Biggl(
		\prod_{V\in \vres} \int_{0}^{1} \pi_{z_{V}}(t_{V})\,\der t_{V}
		\Biggr) \Biggl[ \prod_{u\in\th}
		\frac{\nu_{u}^{m_{u}+1}}{m_{u}!} \Biggr] \cdot \\
	&\quad \cdot 
		\Biggl(\prod_{\ell\in L_{0}}
		g_{n_{\ell}}(\nu_{\ell}(\mathbf{t}))\Biggl)
	 	\Biggl( \prod_{\ell\in L_{1}} \om\nu_{\ell_m(\ell)}
		\frac{\p}{\p\mu_{m}}g_{n_{\ell}}(\nu_{\ell}(\mathbf{t}))
		\Biggl)\cdot \\
	&\quad \cdot 
		\Biggl( \prod_{\ell\in L_{2}} \om\nu_{\ell_{m}(\ell)}
		\om\nu_{\ell_{m'}(\ell)}
		\frac{\p^{2}}{\p\mu_{m}\p\mu_{m'}}
		g_{n_{\ell}}(\nu_{\ell}(\mathbf{t}))\Biggl).
	\end{split}
	\label{eq:s4.renormalization}
\end{equation}

\begin{rem}\label{rem:s4.rem3}
Note that no propagator is derived more than twice: this fact is 
essential for our proof since we have no control on the growth rate 
of the derivatives of the compact support functions 
\eqref{eq:s2.chindef}. 
\end{rem}

After the renormalization procedure has been applied
for all resonances, one check that the
momenta of the lines in $\th$ have changed, with respect to
the original tree $\th_0$ with nonvanishing value, in such a way that
the bound \eqref{eq:s2.distest1elle} still hold.

\begin{lem}\label{lem:s4.pseudo}
	Consider a renormalized tree $\th\in\ti_{\nu,k}^*$,
	obtained from $\th\in\ti_{\nu,k}$ by the iterative
	replacements, described above, that take
	place each time a resonance appears. Then
	the lines of $\th$ inherit the scales of the conjugate
	lines of $\th_0$ and lemma
	\ref{lem:s2.count} applies to $\th$.
\end{lem}

\begin{proof}
The first assertion follows by construction. The second one
can be seen by induction on the generation of the resonances,
by taking into account that for the first generation resonances
the result has been already proved in sect. \ref{sect:canc1}.
So let us suppose that \eqref{eq:s2.distest1}
holds for resonances of any generation $j'$, with $j'<j$.
Consider a line $\ell$ contained inside a resonance
$V\in\vres_{j}$ and outside all resonances in $\vres_{j+1}$
contained inside $V$: then there will be $j$ resonances
$V\equiv W_1 \subset \ldots \subset W_{j}$ containing $\ell$.
Each renormalization produces a change
on the momentum flowing through the line $\ell$, such that,
if $\tilde\nu_{\ell}$ is the momentum flowing through the line
$\ell$ in $\th_0$ and $\nu_{\ell}$ is the momentum
flowing through the conjugate line $\ell$ in $\th$, then
\begin{equation}
	\frac{1}{96q_{n_{\ell}+1}} -
	\sum_{i=1}^{j} \frac{1}{4q_{n_{W_{i}}^{R}}} \leq
	||\om\tilde\nu_{\ell}||\leq \frac{1}{48q_{n_{\ell}}}
	+ \sum_{i=1}^{j} \frac{1}{4q_{n_{W_{i}}^{R}}} .
	\label{eq:s5.a}
\end{equation}
Call $\th^{V}_{0}\in\effe_{\vres_{j}}(\th_0)$ the tree containing $V$
(which is not, in general, the originary tree $\th_0$) and $\th^{V}$ the
tree in $\effe_{V}(\th^{V}_{0})$ obtained by the
action of the group $\pii_{V}$.
As \eqref{eq:s2.distest1elle} is supposed to hold
before renormalizing $V$, for all lines $\ell_{m}$,
$m=1,\ldots,m_{V}$, entering $V$ one has
$||\om\nu_{\ell_{m}}||<1/8q_{n_{\ell_{m}}}$, so that,
by reasoning as in sect. \ref{sect:canc1} to prove
lemma \ref{lem:s3.cambio}, we can conclude that
\begin{equation}
	\bigl| ||\om\nu_{\ell}|| - ||\om\tilde\nu_{\ell}|| \bigr|
	\leq \frac{1}{4q_{n_{V}^{R}}} , \qquad ||\om\nu_{\ell}||\geq
	\frac{1}{4q_{n_{V}^{R}}} , \qquad ||\om\tilde\nu_{\ell}||\geq 
	\frac{1}{4q_{n_{V}^{R}}} ,
	\label{eq:s5.b}
\end{equation}
where $\nu_{\ell}$ is the momentum flowing through
the line $\ell$ in $\th^V$.

In order that $\ell$ be contained inside $V=W_1$, one must have
$1/48q_{n_{\ell}}\geq 1/4q_{n_{V}^{R}}$; moreover if $j_1 =
\lfloor(j-1)/2\rfloor$ and $j_2=\lfloor j/2 \rfloor$
(here $\lfloor \cdot\rfloor$ denotes the integer part), one has
\begin{equation}
	q_{n_{W_{1}}} \leq \frac{q_{n_{W_{3}}}}{2} \leq \ldots
	\leq \frac{q_{n_{W_{j_{1}}}}}{2^{j_{1}}} , \qquad
	q_{n_{W_{2}}} \leq \frac{q_{n_{W_{4}}}}{2} \leq \ldots
	\leq \frac{q_{n_{W_{j_{2}}}}}{2^{j_{2}}} ,
	\label{eq:s5.c}
\end{equation}
(simply use that $q_{n+1}\geq q_{n}$ and $q_{n+2}\geq 2q_{n}$
for any $n\geq 0$). Then one can write
\begin{equation}
	||\om\nu_{\ell}||\leq \frac{1}{48q_{n_{\ell}}}
	+ \frac{1}{4q_{n_{V}^{R}}} \Bigl(
	\sum_{i=1}^{j_1} \frac{1}{2^{i}} +
	\sum_{i=1}^{j_2} \frac{1}{2^{i}} +	
	\Bigr) \leq
	\frac{1}{48q_{n_{\ell}}} + \frac{1}{q_{n_{V}^{R}}};
	\label{eq:s5.d}
\end{equation}
this is bounded from above by $5/48q_{n_{\ell}}$.
Likewise one finds
\begin{equation}
	||\om\nu_{\ell}||\geq \frac{1}{96q_{n_{\ell}+1}}
	- \frac{1}{4q_{n_{V}^{R}}} \Bigl(
	\sum_{i=1}^{j_1} \frac{1}{2^{i}} +
	\sum_{i=1}^{j_2} \frac{1}{2^{i}} +	
	\Bigr) \geq \frac{1}{96q_{n_{\ell}+1}}
	- \frac{1}{q_{n_{V}^{R}}};
	\label{eq:s5.e}
\end{equation}
this is bounded from below by $1/192q_{n_{\ell}+1}$ if
$1/96q_{n_{\ell}+1}> 2/q_{n_{V}^{R}}$ and
by $1/768q_{n_{\ell}+1}$ if $1/96q_{n_{\ell}+1}\leq 2/q_{n_{V}^{R}}$.

Then \eqref{eq:s2.distest1} holds also for any line $\ell$
contained inside $V_0$, if $V$ is a resonance in $\vres_{j}$.
As any next renormalization is on resonances $V\in\vres_{j'}$,
with $j'>j$, so that it does not shift the line $\ell$,
the momentum $\nu_{\ell}$ changes no more, so that
the inductive proof is complete.
\end{proof}

Then in \eqref{eq:s4.renormalization} we can bound, for $\ell\in L_1$:
\begin{equation}
	\begin{split}
		& \biggl| \om\nu_{\ell_m(\ell)}
		\frac{\p}{\p\mu_{m}} 
		g_{n_{\ell}}(\nu_{\ell}(\mathbf{t})) \biggr| \leq\\
	& \qquad\qquad \leq D_{9} ||\om\nu_{\ell_m(\ell)}||
		\bigl(768q_{n_{\ell}+1} \bigr)^{3} \\
	& \qquad\qquad \leq D_{9} ||\om\nu_{\ell_{m}(\ell)}||
		\bigl(768q_{n_{\ell}+1} \bigr)^{3} \prod_{i=0}^{p-1}
	\frac{||\om\nu_{\ell^{0}_{W_{i}}}||}{||\om\nu_{\ell^{0}_{W_{i}}}||}\\
	& \qquad\qquad \leq D_{9} \bigl(768q_{n_{\ell}+1}\bigr)^{2}
		\Biggl[\prod_{i=0}^{p}||\om\nu_{\ell^{0}_{W_{i}}}||\Biggr]
		\Biggl[\prod_{i=0}^{p}\bigl(768q_{n_{W_{i}}+1}\bigr)\Biggr],
	\end{split}
	\label{eq:s4.422}
\end{equation}
where $\wres(\ell)=\{W_0,\ldots,W_p\}$ is the simple cloud of $\ell$,
and, for $\ell\in L_2$:
\begin{equation}
	\begin{split}
		& \biggl| \om\nu_{\ell_m(\ell)} \om\nu_{\ell_{m'}(\ell)}
		\frac{\p^2}{\p\mu_{m}\p\mu_{m'}}
		g_{n_{\ell}}(\nu_{\ell}(\mathbf{t})) \biggr|\leq\\
	& \qquad\qquad \leq D_{9} ||\om\nu_{\ell_m(\ell)}||
		\,||\om\nu_{\ell_{m'}(\ell)}||
		\bigl(768q_{n_{\ell}+1} \bigr)^{4} \\
	& \qquad\qquad \leq D_{9} ||\om\nu_{\ell_{m}(\ell)}||
		\,||\om\nu_{\ell_{m'}(\ell)}||
		\bigl(768q_{n_{\ell}+1} \bigr)^{4} \prod_{i=0}^{p-1}
	\frac{||\om\nu_{\ell^{0}_{W_{i}}}||}{||\om\nu_{\ell^{0}_{W_{i}}}||}
		\prod_{i'=0}^{p'-1} \frac{||\om\nu_{\ell^{0}_{W_{i'}}}||}
		{||\om\nu_{\ell^{0}_{W_{i'}}}||} \\
	& \qquad\qquad \leq D_{9} \bigl(768q_{n_{\ell}+1}\bigr)^{2}
		\Biggl[\prod_{i=0}^{p}||\om\nu_{\ell^{0}_{W_{i}}}||\Biggr]
		\Biggl[\prod_{i=0}^{p}\bigl(768q_{n_{W_{i}}+1}\bigr)\Biggr] \\
	& \qquad\qquad \quad \Biggl[\prod_{i'=0}^{p'}
		||\om\nu_{\ell^{0}_{W_{i'}}}||\Biggr]
		\Biggl[\prod_{i'=0}^{p'}\bigl(768q_{n_{W_{i'}}+1}\bigr)\Biggr],
	\end{split}
	\label{eq:s4.423}
\end{equation}
where $\wres_{-}(\ell)=\{W_0,\ldots,W_{p'}\}$ is the minor cloud
and $\wres_{+}(\ell)=\{W_0,\ldots,W_{p}\}$ is the major cloud of $\ell$.

Note that \eqref{eq:s4.422} and \eqref{eq:s4.423} give a factor
\begin{equation}
	||\om\nu_{\ell^{0}_{W_{i}}}|| \bigl(768q_{n_{W_{i}}+1}\bigr)
	\label{eq:s4.gain1}
\end{equation}
for each resonance $W_i$ belonging to the (simple or minor or major)
cloud of $\ell$. As each resonance belongs to the cloud of
some line internal to it and each resonance contains two derived
lines or one line derived twice (by definition of the
renormalization procedure), then one concludes that a factor equal
to the square of \eqref{eq:s4.gain1} is obtained for each resonance.

If we note that each underived propagator can be bounded again using 
\eqref{eq:s3.charfunest2} with $p=0$, then we can summarize the 
bounds \eqref{eq:s4.422} $\div$ \eqref{eq:s4.423} stating that, for 
each resummed tree $\th$, we have:
\begin{itemize}
	\item  for each resonance $V$, a factor
	$||\om\nu_{\ell^{0}_{V}}||^{2}$ times a factor 
	$(768q_{n_{V}+1})^{2}$;

	\item  for each line $\ell$, a factor 
	$D_{9}(768q_{n_{\ell}+1})^{2}$ (as the factors 
	$(768q_{n_{\ell}+1})^{p}$, $p=1, 2$, appearing when the 
	corresponding propagator is derived, are taken into account by the 
	factors associated to the resonances, see the item above);
\end{itemize}

Then the statement concerning \eqref{eq:s4.renresbound} is proved.

Once the single summand in \eqref{eq:s4.renormalization} has been
bounded, one is left with the problem of bounding the
number of terms on which the sum is performed.

For each first generation resonance $V$ at most $m_{V}^{2}$ times 
$k_{V}^{2}$ summands are generated by the renormalization procedure 
(see \eqref{eq:s3.lem5}). In general, for each (renormalized) 
resonance, we have to sum over the entering lines
$\{\ell^{V}_{m},\ell^{V}_{m'}\}^{*}$ (corresponding to 
the quantities $\mu_{m}$, $m=1,\dots,m_{V}$, in terms of which the 
renormalized resonance factor is considered a function) and over the 
internal lines $\{\ell_{V}^1,\ell_{V}^{2}\}$
(corresponding to the factors on which the derivatives act).
An estimates on the number of summands generated by the 
renormalization procedure can be obtained by using the
counting lemma \ref{lem:s4.lemma7}.

If $V \in \vres_{j}$, $j \geq 1$, let $\enne_{V}$ be the number of 
$(j+1)$-th generation resonances contained inside $V$. Recall
that $V_{0}$ is the set of lines internal to $V$ which are outside any 
resonance contained in $V$, and denote by $k_{V_{0}}$ the number
of elements in $V_{0}$. 

The renormalization procedure, for each renormalized resonance, 
generates a single or double sum over the entering lines whose momenta 
appear in the quantities $\om\nu_{\ell_{1}}(\mathbf{t})$, \dots, 
$\om\nu_{\ell_{m_{V}}}(\mathbf{t})$, in terms of which the 
resonance factor is expanded: the sum is single if the localization is 
to first order and double if the localization is to second order (see 
\eqref{eq:s4.grendef1} and \eqref{eq:s4.grendef2}). 

Then we find, using lemma \ref{lem:s4.lemma7}, that in the 
renormalization procedure each sum over the entering lines of a first 
generation resonance $V$ is on $m_{V}$ terms, each sum over the 
entering lines of all second order resonances $V' \subset V$ is on 
$k_{V_{0}} + \enne_{V}$ terms, each sum over the entering lines of 
all third generation resonances $V'' \subset V' \subset V$ is on 
$k_{V'_{0}} + \enne_{V'}$, and so on; in general, each sum over the 
entering lines of all the resonances $V' \in \vres_{j+1}$ contained 
inside a resonance $V \in \vres_{j}$ is bounded by $k_{V_{0}} + \enne_{V}$.

Once all generations of resonances have been considered, the overall 
number of summands generated by the renormalization procedure -- by 
taking also into account the sum over the derived lines and using 
remark \ref{rem:s4.rem3} -- is bounded by:
\begin{equation}
	\biggl[\prod_{V\in\vres_{1}}k_{V}^{2}\biggr]
	\biggl[\Bigl(\prod_{V\in\vres_{1}}m_{V}^{2}\Bigr)
	\Bigl(\prod_{V\in\vres}(k_{V_{0}}+\enne_{V})^{2}\Bigr)\biggr]
	\leq e^{6k},
	\label{eq:s4.427}
\end{equation}
where $k$ is the order of the tree $\th$. In fact, just use $x \leq 
e^{x}$
and the obvious inequalities:
\begin{equation}
	\begin{split}
		&\sum_{V\in\vres_{1}}k_{V} \leq k, \\
		&\sum_{V\in\vres_{1}}m_{V} + \sum_{V\in\vres}k_{V_{0}} \leq k, \\
		&\sum_{V\in\vres}\enne_{V} \leq k.
	\end{split}
	\label{eq:428}
\end{equation}
Then the statement after \eqref{eq:s4.renresbound} is proved and the 
constant $D_{3}$ is $e^{6}$. 

Finally one has to count the number of trees. The bound given
in sect. \ref{sect:trees} is no more valid, as a line $\ell\in\th$
can have more than two scale labels. However lemma \ref{lem:s2.scale}
proves that to each line at most $D_{10}=17$ scale labels can be
associated, so that the number of trees in $\ti_{\nu,k}^*$
is bounded by $2^{3k}D_{10}^k$.
Then the bound \eqref{eq:s2.trivbound3} follows, with $D_{4} = 
2^{3}D_{3}D_{9}D_{10}$: this concludes the proof of the theorem.

\vspace{1cm}

\section{Proof of lemma \ref{lem:s2.count}}
\label{sect:proof4}

\noindent
We shall prove inductively on the order $k$ the following bounds:
\begin{subequations}\label{eq:s5.1}
\begin{align}
	&M_{n}(\th) = 0, & \quad &\text{if $k<q_{n}$,}\label{eq:s5.1a}\\
	&M_{n}(\th) \leq \frac{2k}{q_{n}} - 1 + N^{R}_{n}(\th), &
		&\text{if $k \geq q_{n}$,}\label{eq:s5.1b}
\end{align}
\end{subequations}
for any $n \geq 0$, and:
\begin{subequations}\label{eq:s5.2}
\begin{align}
	&M_{n}(\th) = 0, & \quad &\text{if $k < q_{n}$,}\label{eq:s5.2a}\\
	&M_{n}(\th) \leq \frac{k}{q_{n}} + N^{R}_{n}(\th), &
		&\text{if $q_{n}\leq k < \frac{q_{n+1}}{4}$,}\label{eq:s5.2b}\\
	&M_{n}(\th) \leq \frac{k}{q_{n}} + \frac{8k}{q_{n+1}} - 1 +
		N^{R}_{n}(\th), &
		&\text{if $k \geq \frac{q_{n+1}}{4}$,}\label{eq:s5.2c}
\end{align}
\end{subequations}
for $q_{n+1}>4q_{n}$, where $k$ is the order of the tree $\th$. 

Note that \eqref{eq:s5.1a} and \eqref{eq:s5.2a} are simply a 
consequence of lemma \ref{lem:s2.smalltree} of sect. 
\ref{sect:trees}, so we have to prove only \eqref{eq:s5.1b}, 
\eqref{eq:s5.2b} and \eqref{eq:s5.2c}. 

\begin{rem}
If we were only interested in proving the analyticity of the 
invariant curves for rotation numbers satisfying the Bryuno 
condition, then equations \eqref{eq:s5.1} would be sufficient
-- as it would be easy to check by proceeding along the
lines of sect. \ref{sect:canc1} and \ref{sect:canc3}. 
However, in order to find the optimal dependence of the radius of 
convergence $\rho(\om)$ on $\om$, which is the main focus of this 
paper, the more refined bounds $\eqref{eq:s5.2}$ are necessary. 
\end{rem}

\begin{rem}
The proof of \eqref{eq:s5.1} is easier, as it is obvious since it is a 
weaker result.  After dealing with \eqref{eq:s5.2}, the proof of 
\eqref{eq:s5.1} could be left as an exercise: we shall prove it 
explicitely for completeness, and as it could be read as an 
introduction to the more involved proof of \eqref{eq:s5.2}. 
\end{rem}

We shall prove first \eqref{eq:s5.2} (case $q_{n+1}>4q_{n}$) in cases 
\textup{[1]} $\div$ \textup{[3]} below, then \eqref{eq:s5.1} in items 
\textup{[4]} $\div$ \textup{[6]} below.  We proceed by induction, and 
assuming that \eqref{eq:s5.1}, \eqref{eq:s5.2} hold for any $k'<k$ we 
shall show that they hold for $k$ also; their validity for $k=1$ being 
trivial, lemma \ref{lem:s2.count} is proved.  Recall also remark 
\ref{rem:s2.remark5} in sect. \ref{sect:trees} about the way of counting
the resonances on scale $n$ and the resonances with resonance-scale $n$.

$\bullet$ So consider first $q_{n+1}>4q_{n}$. 

\begin{case}[1]
If the root line $\ell$ of $\th$ has scale $\neq n$ and it is not the 
exiting line of a resonance on scale $n$, let us denote with 
$\ell_{1}$, \dots, $\ell_{m}$ the lines entering the last node $u_{0}$ 
of $\th$ and $\th_{1}$, \dots, $\th_{m}$ the subtrees of $\th$ whose 
root lines are those lines.  By construction $M_{n}(\th) = 
M_{n}(\th_{1}) + \dots + M_{n}(\th_{m})$ and $N^{R}_{n}(\th) = 
N^{R}_{n}(\th_{1}) + \dots + N^{R}_{n}(\th_{m})$: the bounds 
\eqref{eq:s5.2} follow inductively by noting that for $k \geq 
q_{n+1}/4$ one has $8k/q_{n+1} - 1 \geq 1$. 
\end{case}

\begin{case}[2]
If the root line $\ell$ of $\th$ has scale $n$, then we can reason as 
follows. Let us denote with $\ell_{1}$, \dots, $\ell_{m}$ the lines on 
scale $\geq n$ which are the nearest to the root line of 
$\th$,\footnote{That is, such that no other line along the paths 
connecting the lines $\ell_{1}$, \dots, $\ell_{m}$ to the root line is 
on scale $\geq n$.} and let $\th_{1}$, \dots, $\th_{m}$ be the 
subtrees with root lines $\ell_{1}$, \dots, $\ell_{m}$. If $m=0$ then 
\eqref{eq:s5.2} follow immediately from lemma \ref{lem:s2.smalltree} 
of sect. \ref{sect:trees}; so let us suppose that $m \geq 1$. Then 
the lines $\ell_{1}$, \dots, $\ell_{m}$ are the entering lines of a 
cluster $T$ (which can degenerate to a single point) having the root 
line of $\th$ as the exiting line. As $\ell$ cannot be the exiting 
line of a resonance on scale $n$, one has:
\begin{equation}
	M_{n}(\th) = 1 + M_{n}(\th_{1}) + \dots + M_{n}(\th_{m}).
	\label{eq:s5.3}
\end{equation}
In general $\tilde{m}$ subtrees among the $m$ considered have orders 
$\geq q_{n+1}/4$, with $0 \leq \tilde{m} \leq m$, while the remaining 
$m_{0} = m - \tilde{m}$ have orders $<q_{n+1}/4$. Let us numerate the 
subtrees so that the first $\tilde{m}$ have orders $\geq q_{n+1}/4$. 

Let us distinguish the cases $k<q_{n+1}/4$ and $k\geq q_{n+1}/4$. 

\begin{case}[2.1]
If $k<q_{n+1}/4$, then $\tilde{m}=0$ and each line entering $T$, by 
lemma \ref{lem:s2.davie} of sect. \ref{sect:trees}, has a momentum 
which is a multiple of $q_{n}$ and, by lemma \ref{lem:s2.smalltree}, 
has a scale label $n$. Therefore the momentum flowing through 
the root line is $\nu = \nu_{T} + s_{0}q_{n}$, for some $s_{0} \in 
\fZ$, with:
\begin{equation}
	\nu_{T} \equiv \sum_{u\in T}\nu_{u}.
	\label{eq:s5.4}
\end{equation}
Moreover also the root line of $\th$ has scale $n$, by assumption, 
and momentum $\nu = sq_{n}$ for some $s\in\fZ$, by lemma 
\ref{lem:s2.davie}, so that $\nu_{T} = (s-s_{0})q_{n} = s'q_{n}$, for 
some integer $s'$.

\begin{case}[2.1.1]
If $s'\neq 0$, then $k_{T}\geq|\nu_{T}|\geq q_{n}$, giving:
\begin{multline}
	M_{n}(\th) \leq 1 + \frac{k_{1}+\dots+k_{m}}{q_{n}} + 
	N^{R}_{n}(\th_{1})+ 
	\dots + N^{R}_{n}(\th_{m}) \leq \\
	1 + \frac{k-k_{T}}{q_{n}} + N^{R}_{n}(\th) \leq
	\frac{k}{q_{n}} + N^{R}_{n}(\th),
	\label{eq:s5.5}
\end{multline}
as $N^{R}_{n}(\th) = N^{R}_{n}(\th_{1})+ \dots + N^{R}_{n}(\th_{m})$, 
and \eqref{eq:s5.2b} follows. 
\end{case}

\begin{case}[2.1.2]
If $s'=0$ and $k_T\geq q_n$, one can reason as in case \textup{[2.1.1]}.
\end{case}

\begin{case}[2.1.3]
If $s'=0$ and $k_T<q_n$, then $T$ is a
resonance with resonance-scale $n$, and:
\begin{multline}
	M_{n}(\th) \leq 1 + \frac{k_{1}+\dots+k_{m}}{q_{n}} + 
	N^{R}_{n}(\th_{1}) + \dots + N^{R}_{n}(\th_{m}) \leq \\
	\leq 1 + \frac{k}{q_{n}} + N^{R}_{n}(\th_{1}) + \dots + 
	N^{R}_{n}(\th_{m}) \leq \frac{k}{q_{n}} + N^{R}_{n}(\th),
	\label{eq:s5.6}
\end{multline}
as $N^{R}_{n}(\th) = 1 + N^{R}_{n}(\th_{1})+ \dots + 
N^{R}_{n}(\th_{m})$, and again \eqref{eq:s5.2b} follows. 
\end{case}

\end{case}

\begin{case}[2.2]
If $k \geq q_{n+1}/4$, assume again inductively the bounds 
\eqref{eq:s5.2}. From \eqref{eq:s5.3} we have:
\begin{equation}
	M_{n}(\th) \leq 1 + \sum_{j=1}^{\tilde{m}}
	\Bigl(\frac{k_{j}}{q_{n}} + \frac{8k_{j}}{q_{n+1}} - 1\Bigr) + 
	\sum_{j=\tilde{m}+1}^{m}\frac{k_{j}}{q_{n}} + 
	\sum_{j=1}^{m}N^{R}_{n}(\th_{j}),
	\label{eq:s5.7}
\end{equation}
where $k_{j}$ is the order of the subtree $\th_{j}$, $j=1,\dots,m$. 

\begin{case}[2.2.1]
If $\tilde{m} \geq 2$, then \eqref{eq:s5.2c} follows immediately. 
\end{case}

\begin{case}[2.2.2]
If $\tilde{m} = 0$, then \eqref{eq:s5.7} gives:
\begin{multline}
	M_{n}(\th) \leq 1 + \frac{k_{1}+\dots+k_{m}}{q_{n}} + 
	\sum_{j=1}^{m}N^{R}_{n}(\th_{j}) \leq 
	1 + \frac{k}{q_{n}} + \sum_{j=1}^{m}N^{R}_{n}(\th_{j}) \leq \\
	\leq \frac{8k}{q_{n+1}} -1 + \frac{k}{q_{n}} + N^{R}_{n}(\th),
	\label{eq:s5.8}
\end{multline}
as we are considering $k$ such that $1 \leq 8k/q_{n+1}-1$ and 
$N^{R}_{n}(\th_{1}) + \dots + N^{R}_{n}(\th_{m}) = N^{R}_{n}(\th)$. 
\end{case}

\begin{case}[2.2.3]
If $\tilde{m} = 1$, then \eqref{eq:s5.7} gives:
\begin{multline}
	M_{n}(\th) \leq 1 + 
	\Bigl(\frac{k_{1}}{q_{n}} + \frac{8k_{1}}{q_{n+1}} - 1\Bigr) + 
	\sum_{j=2}^{m}\frac{k_{j}}{q_{n}} + 
	\sum_{j=1}^{m}N^{R}_{n}(\th_{j}) = \\
	= \frac{k_{1}}{q_{n}} + \frac{8k_{1}}{q_{n+1}} + 
	\frac{k_{0}}{q_{n}} + 
	\sum_{j=1}^{m}N^{R}_{n}(\th_{j}),
	\label{eq:s5.9}
\end{multline}
where $k_{0} = k_{2} + \dots + k_{m}$.

\begin{case}[2.2.3.1]
If in such case $k_{0} \geq q_{n+1}/8$, then we can bound in 
\eqref{eq:s5.9}:
\begin{equation}
	\frac{k_{1}}{q_{n}} + \frac{8k_{1}}{q_{n+1}} + \frac{k_{0}}{q_{n}} 
	\leq \frac{k_{1}+k_{0}}{q_{n}} + \frac{8(k_{1}+k_{0})}{q_{n+1}} - 
	\frac{8k_{0}}{q_{n+1}} \leq \frac{k}{q_{n}} + \frac{8k}{q_{n+1}} -1,
	\label{eq:s5.10}
\end{equation}
and $N^{R}_{n}(\th_{1} + \dots + N^{R}_{n}(\th_{m}) = 
N^{R}_{n}(\th)$, so that \eqref{eq:s5.2c} follows. 
\end{case}

\begin{case}[2.2.3.2]
If $k_{0} < q_{n+1}/8$, then, denoting with $\nu$ and $\nu_{1}$ the 
momenta flowing through the root line $\ell$ of $\th$ and the root line 
$\ell_{1}$ of $\th_{1}$ respectively, one has:
\begin{equation}
	||\om(\nu-\nu_{1})|| \leq ||\om\nu|| + ||\om\nu_{1}|| \leq 
	\frac{1}{4q_{n}},
	\label{eq:s5.11}
\end{equation}
as both $\ell$ and $\ell_{1}$ are on scale $\geq n$
(see remark \ref{rem:s2.rem2} in sect. \ref{sect:trees} and use
\eqref{eq:s2.distest1}).  Then either 
$|\nu-\nu_{1}| \geq q_{n+1}/4$ or $\nu-\nu_{1}=\tilde{s}q_{n}$, 
$\tilde{s}\in\fZ$, by lemma \ref{lem:s2.davie} of sect.  
\ref{sect:trees}. 

\begin{case}[2.2.3.2.1]
If $|\nu-\nu_{1}| \geq q_{n+1}/4$, noting that 
$\nu=\nu_{1}+\nu_{T}+\nu_{0}$, where $\nu_{0} = s_{0}q_{n}$ (with 
$s_{0}\in\fZ$ and $|\nu_{0}| \leq k_{0} < q_{n+1}/8$) is the sum of 
the momenta flowing through the root lines of the $m_{0}$ subtrees 
entering $T$ with orders $<q_{n+1}/4$ and $\nu_{T}$ is defined by 
\eqref{eq:s5.4}, one has:
\begin{equation}
	k_{T} \geq |\nu_{T}| \geq |\nu-\nu_{1}| - |\nu_{0}| 
	\geq \frac{q_{n+1}}{8},
	\label{eq:s5.12}
\end{equation}
so that in \eqref{eq:s5.9} one can bound:
\begin{multline}
	\frac{k_{1}}{q_{n}} + \frac{8k_{1}}{q_{n+1}} + \frac{k_{0}}{q_{n}} 
	\leq \frac{k-k_{T}}{q_{n}} + \frac{8(k-k_{0}-k_{T})}{q_{n+1}} \leq 
	\frac{k}{q_{n}} + \frac{8(k-k_{T})}{q_{n+1}} \leq \\
	\leq \frac{k}{q_{n}} + \frac{8k}{q_{n+1}} - 1,
	\label{eq:s5.13}
\end{multline}
and $N^{R}_{n}(\th_{1}) + \dots + N^{R}_{n}(\th_{m}) = 
N^{R}_{n}(\th)$, so that \eqref{eq:s5.2c} follows again. 
\end{case}

\begin{case}[2.2.3.2.2]
If $\nu-\nu_{1} = \tilde{s}q_{n}$, $\tilde{s}\in\fZ$, then:
\begin{equation}
	\nu_{T} = \nu-\nu_{1}-\nu_{0} = (\tilde{s}-s_{0}) \equiv sq_{n},
	\label{eq:s5.14}
\end{equation}
where $s\in\fZ$.

\begin{case}[2.2.3.2.2.1]
If $s \neq 0$, then $k_{T} \geq q_{n}$, so that in \eqref{eq:s5.3} one
has:
\begin{equation}
	\frac{k_{1}}{q_{n}} + \frac{8k_{1}}{q_{n+1}} + \frac{k_{0}}{q_{n}} 
	\leq \frac{k-k_{T}}{q_{n}} - \frac{8k}{q_{n+1}} 
	\leq \frac{k}{q_{n}} - 1 + \frac{8k}{q_{n+1}},
	\label{eq:s5.15}
\end{equation}
and $N^{R}_{n}(\th_{1}) + \dots + N^{R}_{n}(\th_{m}) = 
N^{R}_{n}(\th)$, so implying \eqref{eq:s5.2c}.
\end{case}

\begin{case}[2.2.3.2.2.2]
If $s=0$ (\ie $\nu_{T}=0$) and $k_T\geq q_n$, one can proceed
as in case \textup{[2.2.3.2.2.1]}.
\end{case}

\begin{case}[2.2.3.2.2.3]
If $s=0$ and $k_{T} < q_{n}$, then $T$ is a resonance with resonance-scale
$n$,\footnote{If $m_{0}=0$, then $n\equiv\nu_{\ell} = \nu_{\ell_{1}}$ 
so that $n_{\ell} \leq n_{\ell_{1}} \leq n_{\ell} + 1$, by 
construction and by remark \ref{rem:s2.rem2}.} so that $N^{R}_{n}(\th) 
= 1 + N^{R}_{n}(\th_{1}) + \dots + N^{R}_{n}(\th_{m})$, hence
\eqref{eq:s5.9} gives:
\begin{equation}
	M_{n}(\th) \leq \frac{k}{q_{n}} + \frac{8k}{q_{n+1}} - 1
	+ 1+\sum_{j=1}^{m}N^{R}_{n}(\th_{j}) \leq 
	\frac{k}{q_{n}} + \frac{8k}{q_{n+1}} - 1 + N^{R}_{n}(\th),
	\label{eq:s5.16}
\end{equation}
and \eqref{eq:s5.2c} follows. 
\end{case}
\end{case}
\end{case}
\end{case}
\end{case}
\end{case}

\begin{case}[3]
If the root line $\ell$ of $\th$ is on scale $>n$ and it is the 
exiting line of a resonance $V_{n}$ on scale $n$, let us denote with 
$\ell_{1}$, \dots, $\ell_{m}$ the lines on scale $\geq n$ which are 
the nearest to the root line of $\th$, and let $\th_{1}$, \dots, 
$\th_{m}$ be the subtrees with root lines $\ell_{1}$, \dots, 
$\ell_{m}$; some of these lines -- at least one -- are lines on scale 
$n$ inside $V_{n}$.\footnote{Otherwise $V_{n}$ would not contain any 
line on scale $n$, so that it would not be a resonance on scale $n$ 
as we are supposing.} Let $T$ be the cluster which the lines 
$\ell_{1}$, \dots, $\ell_{m}$ enter; of course $T \subset V_{n}$ and $T$ 
can degenerate into a single point. As in case \textup{[2]}, let 
$\tilde{m}$ be the number of subtrees among the $m$ considered which 
have orders $\geq q_{n+1}/4$, and again let us numerate the subtrees 
in such a way that the ones with orders $\geq q_{n+1}/4$ are the 
first $\tilde{m}$. 

Note that $k\geq q_{n+1}$ (otherwise $\ell$ could not be on scale
$>n$) and
\begin{equation}
	M_{n}(\th) = 1 + M_{n}(\th_{1}) + \dots + M_{n}(\th_{m}),
	\label{eq:s5.17}
\end{equation}
as the root line $\ell$ contributes one unit to $P_{n}(\th)$ and does 
not contribute to $N_{n}(\th)$. Note also that if $T$ is a resonance 
then its resonance scale is $n$.

\begin{case}[3.1]
If $T$ is not a resonance, then:
\begin{equation}
	N^{R}_{n}(\th) = N^{R}_{n}(\th_{1}) + \dots + N^{R}_{n}(\th_{m}).
	\label{eq:s5.18}
\end{equation}
By induction \eqref{eq:s5.2} and \eqref{eq:s5.17} imply:
\begin{equation}
	M_{n}(\th) \leq 1 + \sum_{j=1}^{\tilde{m}}
	\biggl(\frac{k_{j}}{q_{n}} + \frac{8k_{j}}{q_{n+1}} - 1\biggr) + 
	\sum_{j=1}^{m}\frac{k_{j}}{q_{n}} + \sum_{j=1}^{m}N^{R}_{n}(\th_{j}),
	\label{eq:s5.19}
\end{equation}
where $k_{j}$ are the orders of the subtrees $\th_{j}$, $j=1, \dots, m$. 

\begin{case}[3.1.1]
If $\tilde{m}=2$, then \eqref{eq:s5.2c} follows immediately.
\end{case}

\begin{case}[3.1.2]
The case $\tilde{m}=0$ is impossible because $T$ is contained inside 
a resonance $V_{n}$ on scale $n$, so that at least one of the 
subtrees entering $T$ must have order $\geq q_{n+1}/4$ -- otherwise no 
line on scale $>n$ could enter $V_{n}$, see lemma 
\ref{lem:s2.smalltree}. 
\end{case}

\begin{case}[3.1.3]
If $\tilde{m}=1$ let $k_{0} = k_{2} + \dots + k_{m}$; then the case 
$k_{0} \geq q_{n+1}/8$ can be dealt with as in case 
\textup{[2.2.3.1]}; if $k_{0} < q_{n+1}/8$, we deduce from lemma 
\ref{lem:s2.davie} that either $|\nu-\nu_{1}| \geq q_{n+1}/4$ or 
$\nu-\nu_{1}=\tilde{s}q_{n}$, using the same notations of case 
\textup{[2.2.3.2]}.

The first case can be discussed as in case \textup{[2.2.3.2.1]}, while in 
the second case we find, as in case \textup{[2.2.3.2.2]}, that $\nu_{T} = 
\nu - \nu_{1} - \nu_{0} = sq_{n}$, with either $s \neq 0$ or
$s=0$ and $k_T\geq q_n$
(otherwise $T$ would be a resonance), so that the conclusions in 
cases \textup{[2.2.3.2.2.1]} and \textup{[2.2.3.2.2.2]}
can be inherited in the present case and \eqref{eq:s5.2c} follows again.
\end{case}
\end{case}

\begin{case}[3.2]
If $T$ is a resonance, then its resonance-scale is $n$ (and
all its entering lines are on scale $n$; see item
\ref{enum:s2.res1} in the definition of resonance), so that:
\begin{equation}
	N^{R}_{n}(\th) = 1 + N^{R}_{n}(\th_{1}) + \dots + N^{R}_{n}(\th_{m}).
	\label{eq:s5.20}
\end{equation}
The discussion goes on as in case \textup{[3.1]} above, with the only 
difference that now, when $\tilde{m} = 1$ (and $k_T<q_n$,
$k_{0} < q_{n+1}/8$), the 
case $\nu_{T} = 0$ (\ie $\nu_{T} = sq_{n}$, with $s=0$) is the only 
possible since $T$ is a resonance. In such a case:
\begin{equation}
	M_{n}(\th) \leq 1 + \frac{k_1}{q_{n}} +
	\frac{8k_1}{q_{n+1}} - 1 + \frac{k_0}{q_{n}} +
	\sum_{j=1}^{m}N^{R}_{n}(\th_{j}) \leq 
	\frac{k}{q_{n}} + \frac{8k}{q_{n+1}} - 1 + N^{R}_{n}(\th),
	\label{eq:s5.21}
\end{equation}
and \eqref{eq:s5.2c} follows once more. 
\end{case}
\end{case}

$\bullet$ Now we prove \eqref{eq:s5.1}. 

\begin{case}[4]
If the root line $\ell$ of $\th$ as scale $\neq n$ and it is not the 
entering line of a resonance on scale $n$, let us denote with 
$\ell_{1}$, \dots, $\ell_{m}$ the lines entering the last node $u_{0}$ 
of $\th$.  By construction $M_{n}(\th) = M_{n}(\th_{1}) + \dots + 
M_{n}(\th_{m})$ and $N^{R}_{n}(\th) = N^{R}_{n}(\th_{1}) + \dots + 
N^{R}_{n}(\th_{m})$ so that the bound \eqref{eq:s5.1} follows 
immediately by induction. 
\end{case}

\begin{case}[5]
If the root line $\ell$ of $\th$ has scale $n$, using the same 
notations as in case \textup{[2]}, denote with $\ell_{1}$, \dots, 
$\ell_{m}$ the lines on scale $\geq n$ which are nearest to the root 
line of $\th$, and let $\th_{1}$, \dots, $\th_{m}$ be the subtrees 
with these lines as root lines. Then such lines are the entering 
lines of a cluster $T$ (which can degenerate into a single point)
having the root line of $\th$ as the exiting line. We have:
\begin{equation}
	M_{n}(\th) = 1 + M_{n}(\th_{1}) + \dots + M_{n}(\th_{m}).
	\label{eq:s5.22}
\end{equation}
Assuming again inductively the bounds \eqref{eq:s5.1}, from 
\eqref{eq:s5.22} we have:
\begin{equation}
	M_{n}(\th) \leq 1 + 
	\sum_{j=1}^{m}\biggl(\frac{2k_{j}}{q_{n}}-1\biggr) + 
	\sum_{j=1}^{m}N^{R}_{n}(\th_{j}),
	\label{eq:s5.23}
\end{equation}
where $k_{j}$ is the order of the subtree $\th_{j}$, $j=1, \dots, m$. 

\begin{case}[5.1]
If $m \geq 2$, then \eqref{eq:s5.1b} follows immediately.
\end{case}

\begin{case}[5.2]
If $m=0$, then $M_{n}(\th) = 1$. As $\ell$ is on scale $n$, the order 
$k$ of $\th$ has to be $k \geq q_{n}$, so that:
\begin{equation}
	M_{n}(\th) = 1 \leq \frac{2k}{q_{n}} - 1, \quad N^{R}_{n}(\th) = 0,
	\label{eq:s5.24}
\end{equation}
and \eqref{eq:s5.1b} follows again.
\end{case}

\begin{case}[5.3]
If $m=1$, then \eqref{eq:s5.23} gives:
\begin{equation}
	M_{n}(\th) \leq 1 + \biggl(\frac{2k_{1}}{q_{n}} - 1\biggr) + 
	N^{R}_{n}(\th_{1}) = \frac{2k_{1}}{q_{n}} + N^{R}_{n}(\th_{1}).
	\label{eq:s5.25}
\end{equation}
Denoting with $\nu$ and $\nu_{1}$ the momenta flowing, respectively, 
through the root line $\ell$ of $\th$ and through the root line 
$\ell_{1}$ of $\th_{1}$, we have:
\begin{equation}
	||\om(\nu-\nu_{1})|| \leq ||\om\nu|| + ||\om\nu_{1}|| \leq 
	\frac{1}{4q_{n}},
	\label{eq:s5.26}
\end{equation}
as both $\ell$ and $\ell_{1}$ are on scale $\geq n$ (see remark 
\ref{rem:s2.rem2} in page \pageref{rem:s2.rem2} and
use \eqref{eq:s2.distest1}). Then, as $\nu_{T} = 
\nu-\nu_{1}$, either $|\nu_{T}| \geq q_{n}$ or $\nu_{T}=0$. 

\begin{case}[5.3.1]
If $|\nu_{T}| \geq q_{n}$, then $k_{T} \geq |\nu_{T}| \geq q_{n}$ and 
$N^{R}_{n}(\th_{1}) + \dots + N^{R}_{n}(\th_{m}) = N^{R}_{n}(\th)$ 
(since $T$ is not a resonance), so that \eqref{eq:s5.25} gives:
\begin{equation}
	M_{n}(\th) \leq \frac{2k}{q_{n}} - \frac{2k_{T}}{q_{n}} + 
	N^{R}_{n}(\th_{1}) \leq \frac{2k}{q_{n}} - 1 + N^{R}_{n}(\th_{1}),
	\label{eq:s5.27}
\end{equation}
and \eqref{eq:s5.1b} follows.
\end{case}

\begin{case}[5.3.2]
If $\nu_T=0$ and $k_T\geq q_n$, one can reason as in case
\textup{[5.3.1]}.
\end{case}

\begin{case}[5.3.3]
If $\nu_{T}=0$ and $k_T<q_n$,
then $\nu_{1}=\nu$ and either $n_{\ell_{1}}=n$ or 
$n_{\ell_{1}} = n+1$ (see remark \ref{rem:s2.rem1} in page 
\pageref{rem:s2.rem1}): then $T$ is a resonance with resonance scale 
$n$, so that $1 + N^{R}_{n}(\th_{1}) + \dots + N^{R}_{n}(\th_{m}) = 
N^{R}_{n}(\th)$, hence \eqref{eq:s5.25} gives:
\begin{equation}
	M_{n}(\th) \leq \biggl(\frac{2k}{q_{n}} - 1\biggr) + 1 + 
	N^{R}_{n}(\th_{1}) \leq \frac{2k}{q_{n}} - 1 + N^{R}_{n}(\th_{1}),
	\label{eq:s5.28}
\end{equation}
and \eqref{eq:s5.1} follows again.
\end{case}
\end{case}
\end{case}

\begin{case}[6]
If the root line $\ell$ of $\th$ is on scale $>n$ and it is the 
exiting line of a resonance $V_{n}$, as in case \textup{[3]} above, 
denote with $\ell_{1}$, \dots, $\ell_{m}$ the lines on scale $\geq n$ 
wich are nearest to the root line of $\th$, and let $\th_{1}$, \dots, 
$\th_{m}$ be the subtree of $\th$ of which these lines are root lines. 
Some of these lines -- at least one -- are lines on scale $n$ inside 
$V_{n}$. Let $T$ be the cluster which the lines $\ell_{1}$, \dots, 
$\ell_{m}$ enter; of course $T \subset V_{n}$, and $T$ can degenerate 
into a single point. 

Note that as in case \textup{[3]}:
\begin{equation}
	M_{n}(\th) = 1 + M_{n}(\th_{1}) + \dots + M_{n}(\th_{m}),
	\label{eq:s5.29}
\end{equation}
as the root line $\ell$ contributes one unit to $P_{n}(\th)$ and does 
not contribute to $N_{n}(\th)$, and that if $T$ is a resonance then 
its resonance scale is $n$. 

\begin{case}[6.1]
If $T$ is not a resonance, then:
\begin{equation}
	N^{R}_{n}(\th) = N^{R}_{n}(\th_{1}) + \dots + N^{R}_{n}(\th_{m}).
	\label{eq:s5.30}
\end{equation}
By induction, \eqref{eq:s5.1} and \eqref{eq:s5.29} imply:
\begin{equation}
	M_{n}(\th) \leq 1 + \sum_{j=1}^{m}
	\biggl(\frac{2k_{j}}{q_{n}} - 1\biggr) + 
	\sum_{j=1}^{m}N^{R}_{n}(\th_{j}),
	\label{eq:s5.31}
\end{equation}
where $k_{j}$ are the orders of the subtrees $\th_{j}$, $j=1, \dots, m$. 

\begin{case}[6.1.1]
If $m=2$, then \eqref{eq:s5.1b} follows immediately.
\end{case}

\begin{case}[6.1.2]
The case $m=0$ is impossible (see case \textup{[3.1.2]}).
\end{case}

\begin{case}[6.1.3]
If $m=1$ in \eqref{eq:s5.31}, we have $\nu_{T} = \nu-\nu_{1}$, so 
that $|\nu_{T}|\geq q_{n}$ (as $\nu_{T}\neq 0$, otherwise $T$ would be 
a resonance). Then we can go on along the lines of case 
\textup{[5.3.1]} in order to obtain \eqref{eq:s5.1b}. 
\end{case}
\end{case}

\begin{case}[6.2]
If $T$ is a resonance, then its resonance scale is $n$, so that:
\begin{equation}
	N^{R}_{n}(\th) = 1 + N^{R}_{n}(\th_{1}),
	\label{eq:s5.32}
\end{equation}
and the discussion goes on as in case \textup{[6.1]}, with the only 
difference that now, for $m=1$, the case $\nu_{T}=0$ is the only 
possible as $T$ is supposed to be a resonance. In such a case:
\begin{equation}
	M_{n}(\th) \leq 1 + \biggl(\frac{2k}{q_{n}} - 1\biggr) + 
	N^{R}_{n}(\th_{1}) \leq \frac{2k}{q_{n}} - 1 + N^{R}_{n}(\th),
	\label{eq:s5.33}
\end{equation}
implying again \eqref{eq:s5.1b}.
\end{case}
\end{case}

$\bullet$ Finally, to deduce \eqref{eq:s2.count} from \eqref{eq:s5.1}
and \eqref{eq:s5.2}, simply note that, for $q_{n+1} \leq 4q_{n}$, we
have $2k/q_{n} \leq 8k/q_{n+1}$; them lemma \ref{lem:s2.count} follows. 

\begin{rem}\label{rem:s5.rem14}
Note that the correspondence between momenta and scale labels
has been used only through the inequality \eqref{eq:s2.distest1elle}.
As we have seen in sect. \ref{sect:canc3} the renormalization
procedure can shift the ``original'' momenta flowing through the lines
of a bounded quantity which does not alter such an inequality.
This allow us to apply lemma 4 also to the renormalized trees,
as it was repeatedly claimed in the previous sections.
\end{rem}

\vspace{1cm}

\section{Proof of lemma \ref{lem:s4.lemma6}}
\label{sect:proof6}

\noindent
As far as only the localized resonance factor is involved, the momenta 
flowing through the lines entering any resonance are set to zero, so 
that it does not matter if such momenta are interpolated or not (\ie if 
they are of the form $\boldsymbol{\nu}$ or 
$\boldsymbol{\nu}(\mathbf{t})$). In particular, the case of first 
generation resonances (discussed in sect. \ref{sect:canc1}) is 
included in lemma \ref{lem:s4.lemma6}. 

A basic property of the trees belonging to the resonance family 
$\effe_{V}(\th)$ is that the difference between their values is only 
in the resonance factor: for any tree $\th'\in\effe_{V}(\th)$, we can 
write:
\begin{equation}
	\Val(\th') = \aaa(\th)\vu_{V}(\th'),
	\label{eq:s6.1}
\end{equation}
for some factor $\aaa(\th)$ which is the same for all 
$\th'\in\effe_{V}(\th)$. This simply follows from the fact that the 
transformations in $\pii_{V}$ do not touch the part of the tree $\th$ 
which is outside the resonance $V$.
Therefore a cancellation between 
localized resonance factors yields a cancellation between tree 
values (in which the resonance factor has been localized of course). 

By item \ref{enum:s2.res1} in the definition of resonance and by
definition of $V_{0}$, one has
\begin{equation}
	\sum_{u\in V_{0}} \nu_u = 0 ;
	\label{eq:s2.sumres0}
\end{equation}
moreover, given an entering line $\ell_{m}$ of $V$, if
$\ell_{m}\in L_{V}^{R}$ and $\tilde V_{0}=V_0(\ell_{m})$, then
\begin{equation}
	\sum_{u\in \tilde V_{0}} \nu_u \equiv
	\sum_{u\in V_{0}(\ell_m)} \nu_u = 0 .
	\label{eq:s2.sumres1}
\end{equation}
In general we can write, for any tree $\th'\in\effe_{V}(\th)$,
\begin{equation}
	\elle\vu_{V}(\th') = \bbb(\th') \elle\vu_{V_0}(\th')
	\prod_{\ell\in L_{V}^{R}} \elle \vu_{V(\ell)}(\th') ,
	\label{eq:s6.11}
\end{equation}
where $\vu_{V_{0}}(\th')$ and $\vu_{V(\ell)}(\th')$ are defined as
the resonance factor $\vu_{V}(\th')$, but with the product ranging only
over nodes and lines internal to $V_{0}$ and $V(\ell)$, respectively,
while $\elle\vu_{V_{0}}(\th')$ and $\elle\vu_{V(\ell)}(\th')$
are obtained from $\vu_{V_{0}}(\th')$ and $\vu_{V(\ell)}(\th')$,
respectively, by replacing $\nu_{\ell}$ with $\nu_{\ell}^0$ in $V$,
for all lines $\ell\in V$.
In \eqref{eq:s6.11} $\bbb(\th')$ takes into account all other factors
(if there are any), alwyas evaluated with $\nu_{\ell}$ replaced
with $\nu_{\ell}^0$, $\ell\in V$.
Note that, as $\aaa(\th)$ in \eqref{eq:s6.1},
also $\bbb(\th')$ is the same for all $\th'\in\effe_{V}(\th)$,
so that one can set $\bbb(\th')=\bbb(\th)$ and write:
\begin{equation}
	\Val(\th') = \aaa(\th)\vu_{V}(\th'), \qquad
	\elle \vu_{V}(\th') = \bbb(\th) \elle \vu_{V_0}(\th')\prod_{\ell\in
	L_{V}^{R}} \elle \vu_{V(\ell)}(\th') .
	\label{eq:s6.12}
\end{equation}

\begin{case}[1]
If $z_{V}=1$ the localized resonance factor is given by the resonance 
factor computed for $\mu_{1} = \dots = \mu_{m} = 0$.

Summing the localized resonance factors corresponding to the trees 
belonging to $\effe_{V}(\th)$, we can group them into subfamilies of 
inequivalent trees whose contributions are different as for each node 
$u\in V$ there is a factor;
\begin{equation}
	\frac{1}{m_{u}!}\binom{m_{u}}{s_{u}} = \frac{1}{s_{u}!}\frac{1}{r_{u}!},
	\label{eq:s6.2}
\end{equation}
as all terms which are obtained by permutations are summed together 
(this gives the binomial coefficient in the left hand side of the 
above equation), times a factor:
\begin{equation}
	\nu_{u}^{m_{u}+1} = \nu_{u}^{(s_{u}+1)+r_{u}},
	\label{eq:s6.3}
\end{equation}
times a propagator $g_{n_{\ell_{u}}}(\nu^{0}_{\ell_{u}})$ (the last 
factor is missing if corresponding to the line exiting $V$;
see definitions \eqref{eq:s4.lv11}$\div$\eqref{eq:s4.lv13-2}).

Then for $\mu_{1}=\dots=\mu_{m}=0$ we can write:
\begin{equation}
	\begin{split}
		\sum_{\th'\in\effe_{V}(\th)}\elle\vu_{V}(\th') & =
		\sum_{\th'\in\effe_{V}(\th)}
		\Biggl[\prod_{u\in V}\frac{\nu_{u}^{s_{u}+1}}{s_{u}!}\Biggr]
		\Biggl[\prod_{\ell\in V}g_{n_{\ell}}(\nu^{0}_{\ell})\Biggr]
		\cdot \\ 
		& \qquad \cdot 
		\Biggl(\prod_{u\in V_0}\frac{\nu_{u}^{r_{u}}}{r_{u}!}\Biggr)
		\Biggl(\prod_{\ell\in L_{V}^{R}}
		\prod_{u\in V_0(\ell)}
		\frac{\nu_{u}^{r_{u}}}{r_{u}!}\Biggr) = \\
		&  = \Biggl[\prod_{u\in V}
		\frac{\nu_{u}^{s_{u}+1}}{s_{u}!}\Biggr]
		\Biggl[\prod_{\ell\in V}g_{n_{\ell}}(\nu^{0}_{\ell})\Biggr]
		\cdot \\
		& \qquad \cdot \sum_{\th'\in\effe_{V}(\th)}
		\Biggl(\prod_{u\in V_0}\frac{\nu_{u}^{r_{u}}}{r_{u}!}\Biggr)
		\Biggl(\prod_{\ell\in L_{V}^{R}}
		\prod_{u\in V_0(\ell)}\frac{\nu_{u}^{r_{u}}}{r_{u}!}\Biggr),
	\end{split}
	\label{eq:s6.4}
\end{equation}
where we have used the fact that for $\mu_{1}=\dots=\mu_{m}=0$ the 
factors in square brackets have the same value for all 
$\th'\in\effe_{V}(\th)$ (see \eqref{eq:s3.caso2-1} and take into 
account what observed at the beginning of this section). The last sum 
in \eqref{eq:s6.4} can be rewritten as:
\begin{equation}
	\begin{split}
	& \sum_{\th'\in\effe_{V}(\th)}
		\Biggl( \prod_{u\in V_0}
		\frac{\nu_{u}^{r_{u}}}{r_{u}!} \Biggr)
		\Biggl( \prod_{\ell\in L_{V}^{R}} \prod_{u\in V_0(\ell)}
		\frac{\nu_{u}^{r_{u}}}{r_{u}!} \Biggr) = \\
	& \qquad = \Biggl( \sum_{\substack{\{r_{u}\geq 0\}\\ \sum_{u\in V_0}
		r_{u}=m_{V_0}}}
		\prod_{u\in V} \frac{\nu_{u}^{r_{u}}}{r_{u}!} \Biggr)
		\Biggl( \prod_{\tilde V \in \tilde\vres(V) }
		\sum_{\substack{\{r_{u}\geq 0\}\\ \sum_{u\in \tilde V_0}
		r_{u}=1}} \prod_{u\in \tilde V_0}
		\frac{\nu_{u}^{r_{u}}}{r_{u}!} \Biggr) =\\
	& \qquad = \frac{1}{m_{V_0}!} \Biggl(
		\sum_{u\in V_0}\nu_{u}\Biggr)^{m_{V_0}}
		\prod_{\tilde V \in \tilde \vres(V)}
		\Biggl(\sum_{u\in \tilde V_0}\nu_{u}\Biggr) ,
	\label{eq:s6.5}
	\end{split}
\end{equation}
which is zero by definition of resonance (see \eqref{eq:s2.sumres0} and
\eqref{eq:s2.sumres1} above).
\end{case}

\begin{case}[2]
If $z_{V}=2$ the localized resonance factor, with respect to the 
previous case, contains also the first order terms (again computed in 
$\mu_{1}=\dots=\mu_{m}=0$). 

The zero-th order contribution can be discussed as for the case 
$z_{V}=1$, and the same result holds. Also the second order 
contribution vanishes, after summing over the trees 
$\th'\in\effe_{V}(\th)$. To prove this we shall consider separately 
the cases $m_{V}=2$ and $m_{V}=1$. 

In the first case, when the derivative 
$(\p/\p\mu_{m})\vu_{V}(\th;0,\dots,0)$ is considered, let us compare 
all the trees $\th'$ in the subfamily of $\effe_{V}(\th)$ in which 
the line $\ell_{m}$ is kept fixed (call $\bar{u}$ the node which 
such a line enters), while all other lines are shifted (\ie detached 
and reattached to all nodes inside the resonance). The difference with 
respect to the previous case, discussed above, is that the line with 
momentum $\nu_{\ell_{m}}$ can be choosen in $r_{\bar{u}}$ ways among 
the $r_{\bar{u}}$ lines entering the node $\bar{u}\in V$ and outside 
$V$. This means that we can write:
\begin{equation}
	\frac{\nu^{m_{u}+1}_{u}}{m_{u}!}\binom{m_{u}}{s_{u}} = 
	\frac{\nu_{u}^{(s_{u}+1)+r_{u}}}{s_{u}!r_{u}!}
	\label{eq:s6.6}
\end{equation}
for all nodes $u\neq\bar{u}$, and:
\begin{equation}
	\frac{\nu^{m_{\bar{u}}}_{\bar{u}}}{m_{\bar{u}}!}
	\binom{m_{u}}{s_{u}}r_{\bar{u}} = 
	\frac{\nu_{\bar{u}}^{(s_{\bar{u}}+1)+(r_{\bar{u}}-1)}}%
	{s_{\bar{u}}!(r_{\bar{u}}-1)!}
	\label{eq:s6.7}
\end{equation}
for $\bar{u}$. Then we have an expression analogous to 
\eqref{eq:s6.4}, with the only difference that the labels $\{r_{u}\}$ 
have to be replaced with labels $\{r'_{u}\}$, defined as:
\begin{equation}
	 r'_{u}=r_{u}-\delta_{u\bar{u}}, \qquad
	\forall u \text{ either in } V_0 \text{ or in }
	\bigcup_{\tilde V \in \tilde \vres(V)} \tilde V_0 ,
\end{equation}
such that
\begin{equation}
	\sum_{u\in V_0}r'_{u}+\sum_{\tilde V \in \tilde \vres(V)}
	\sum_{u\in \tilde V_0} r'_{u} =m_{V}-1 ;
\end{equation}
so the last sum in the second line of
\eqref{eq:s6.4} has to be replaced by:
\begin{equation}
	\begin{split}
		& \sum_{\th'\in\effe_{V}(\th)}
		\Biggl(\prod_{u\in V_0}\frac{\nu_{u}^{r_{u}}}{r_{u}!}\Biggr)
		\Biggl(\prod_{\ell\in L_{V}^{R}}
		\prod_{u\in V_0(\ell)}\frac{\nu_{u}^{r_{u}}}{r_{u}!}\Biggr) 
		\nu_{\bar{u}} = \\
		& \qquad =
		\Biggl( \sum_{\substack{\{r_{u}\geq 0\}\\ \sum_{u\in V_0}
		r_{u}=m_{V_0}^*}}
		\prod_{u\in V}\frac{\nu_{u}^{r_{u}}}{r_{u}!} \Biggr)
		\Biggl( \prod_{\tilde V \in \tilde\vres(V) }
		\sum_{\substack{\{r_{u}\geq 0\}\\ \sum_{u\in \tilde V_0}
		r_{u}=\zeta^*(\ell)}} \prod_{u\in \tilde V_0}
		\frac{\nu_{u}^{r_{u}}}{r_{u}!} \Biggr) =\\
		& \qquad = \frac{1}{m_{V_0}^*!}
		\Biggl(\sum_{u\in V_0}\nu_{u}\Biggr)^{m_{V_0}^*}
		\prod_{\tilde V \in \tilde \vres(V)}
		\Biggl(\sum_{u\in \tilde V}
		\nu_{u}\Biggr)^{\zeta^*(\tilde V)} ,
	\label{eq:s6.8}
	\end{split}
\end{equation}
where
\begin{equation}
	m_{V_0}^* = \begin{cases}
	m_{V_0} , & \quad \text{ if} \quad \bar{u} \notin V_0 , \\
	m_{V_0}-1 , & \quad \text{ if} \quad \bar{u} \in V_0 , \end{cases}
	\qquad
	\zeta^*(\tilde V) = \begin{cases}
	1 , & \quad \text{ if} \quad \bar{u} \notin \tilde V_0 , \\
	0 , & \quad \text{ if} \quad \bar{u} \in \tilde V_0 , \end{cases}
	\label{eq:s6.ast2}
\end{equation}
so that we have again vanishing contributions (as $m_{V}\geq 2$).

On the contrary, if $m_{V}=1$, the above reasoning does not apply, as 
there is only one entering line.  Anyway the function 
$(\p/\p\mu_{1})\vu_{V}(\th;0)$ is an odd function, as all the 
propagators are even in their arguments, so that the derived 
one\footnote{If $z_{V}=2$, then there is only one derived propagator, 
arising from the renormalization of the resonance $V$ itself.} becomes 
odd, and the numerator contains an even number of $\nu_{u}$'s.  Then 
by reversing the signs of the labels $\nu_{u}$, $u\in V$, the 
numerator will not change, while the overall sign of the denominator 
will change, so that the sum over the first order contributions of 
the localized resonance factors of the two tree values being considered 
vanishes.\footnote{Note that the renormalization transformations of 
type \ref{enum:s3.ren3} are explicitly used in order to implement the 
cancellation mechanism \emph{only} in the case of a resonance $V$ with 
$z_{V}=2$ and $m_{V}=1$. In general not all the transformations
are used for all resonances: in particular, when $z_V=0$,
we consider separately all terms generated by the action
of the group $\pii_{V}$, as there is no need of
additional renormalizations.} 
\end{case}

\begin{case}[3]
Finally if $z_{V}=0$ the localization operator $\elle$ gives zero 
when acting on the resonance factors, so that nothing has to be 
proved. 
\end{case}

\vspace{1cm}

\section{Conclusions}
\label{sect:concl}

\noindent
Our theorem can be related to the result and the methods of 
\cite{BG1}. There we proved that, for $\om \in \fC$, if $\om$ tends to 
a rational number $p/q$ through a path in the complex plane 
non-tangential to the real axis, then the radius of convergence 
satisfies:
\begin{equation}
    \Biggl|\log\rho(\om) + 
    \frac{2}{q}\log\biggl|\om-\frac{p}{q}\biggr|\Biggr| < C_{4}
    \label{eq:s1.bg1}
\end{equation}
for some constant $C_{4}$.

If instead we consider a sequence of \emph{real}, irrational numbers 
tending to a rational value $p/q$, the situation is quite more 
complex. In fact, the limit and its very existence may depend on the 
arithmetic properties of the numbers of the sequence we consider, 
\emph{and on their uniformity in $k$}; namely:
\begin{enumerate}
    \item\label{enum:s1.seqcase1} The sequence $\{\om_{k}\}$ can tend 
    to $p/q$ but, though all the $\om_{k}$ are irrational, some of 
    them are not Bryuno numbers so that for those $B(\om_{k}) = + 
    \infty$ and $\rho(\om_{k}) = 0$.

    \item\label{enum:s1.seqcase2} The sequence $\{\om_{k}\}$ can tend 
    to $p/q$ through Bryuno numbers, or even Diophantine numbers, but 
    they are not uniformly such in $k$ so that $B(\om_{k})$ diverges 
    \emph{faster} than $\log\bigl(|\om_{k}-p/q|^{1/q}\bigr)$ (and so 
    $\rho(\om_{k})$ tends to zero \emph{faster} than 
    $|\om_{k}-p/q|^{2/q}$).  An example can be the sequence of 
    Diophantine (actually even ``noble'') numbers:
\begin{equation}
	\om_{k} = \cfrac{1}{k + \cfrac{1}{2^{k^{2}} + \ga}},
	    \label{eq:s1.noble}
\end{equation}
    where $\ga$ denotes the ``golden mean'':
\begin{equation}
	\ga = \cfrac{1}{1 + \cfrac{1}{1 + \dotsb}} = 
		\frac{\sqrt{5}-1}{2};
    \label{eq:s1.goldenmean}
\end{equation}
    a simple calculation using the recursion relation 
    \eqref{eq:s1.bry} shows that indeed $B(\om_{k}) = O(k)$ while 
    $\om_{k} = O(1/k)$, so that, by taking into account
    also logarithmic corrections in $B(\om_k)$, $\rho(\om_{k}) = 
    O(\om_{k}^{2}e^{-2/\om_{k}})$, that is \emph{much faster} than 
    $\om_{k}^{2}$.

    \item\label{enum:s1.seqcase3} Finally, the sequence $\{\om_{k}\}$ 
    can tend to $p/q$ through a sequence of Bryuno numbers satisfying 
    uniform estimates in $k$, so that an estimate like \eqref{eq:s1.bg1} 
    holds (note that decays slower than $|\om_{k}-p/q|^{2/q}$
    are not possible); an example can be given by the sequence:
\begin{equation}
	\om_{k} = \frac{1}{k + \ga},
	\label{eq:s1.uniform}
\end{equation}
    where again $\ga$ is the golden mean \eqref{eq:s1.goldenmean}. 
\end{enumerate}
Notice that in the numerical calculations of \cite{BM2} only real 
sequences of type \ref{enum:s1.seqcase3} were considered, and that 
sequences of type \ref{enum:s1.seqcase2} are practically inaccessible 
from the numerical point of view. 

Finally, one may ask how much these results can be extended to more 
complicated, and realistic, symplectic maps and continuous time 
Hamiltonian systems. We believe that while some additional 
complications may arise, the really hard  problem (\ie how to handle 
resonances) is already present in the standard map and it was
solved by carefully using the trees formalism and the multiscale
decomposition of the propagators. More general maps and Hamiltonian 
systems, though, as already pointed out in \cite{BG1}, may have 
different, more complicated interpolation properties for the radius of
convergence of their Lindstedt series: this is an area where still much 
work has to be done.

\vspace{1cm}

\end{document}